\documentclass[twocolumn,trackchanges]{aastex6}
\usepackage{amsmath,color,textcomp,url,graphicx,subfigure}
\usepackage{color}
\definecolor{orange}{rgb}{1,0,0} 

\def \MJ {$\,M_{\rm{Jup}}$}

\def \Teff {$T_{\rm{eff}}$}
\def \env {$\sim$\,}
\def \plmo {\,$\pm$\,}
\def \gta {\,\lower 0.5ex\hbox{$\buildrel > \over \sim\ $}\,}   
\def \lta {\,\lower 0.5ex\hbox{$\buildrel < \over \sim\ $}\,}   
\def \tnma {\tablenotemark{a}}
\def \kms {km s$^{-1}$}

\begin{document}
\title{A Search for Photometric Variability in the young T3.5 \\ planetary-mass companion GU Psc \MakeLowercase{b}}
\author{Marie-Eve Naud\altaffilmark{1}*, \'{E}tienne Artigau\altaffilmark{1}, Jason F. Rowe\altaffilmark{1},  Ren\'{e} Doyon\altaffilmark{1}, Lison Malo\altaffilmark{2,1}, Lo\"{i}c Albert\altaffilmark{1}, Jonathan Gagn\'{e}\altaffilmark{3,4}, and Sandie Bouchard\altaffilmark{1}}
\email{*Corresponding author: naud@astro.umontreal.ca}
\affil{\altaffilmark{1} Institut de recherche sur les exoplan\`{e}tes, D\'{e}partement de physique, Universit\'{e} de Montr\'{e}al, Montr\'{e}al, QC H3C 3J7, Canada.\\
\altaffilmark{2} Canada-France-Hawaii Telescope, 65-1238 Mamalahoa Hwy, Kamuela, HI 96743, USA.\\
\altaffilmark{3} Department of Terrestrial Magnetism, Carnegie Institution for Science, 5241 Broad Branch Road NW, Washington, DC 20015, USA.\\
\altaffilmark{4} NASA Sagan Fellow.}

\begin{abstract}
We present a photometric $J$-band variability study of GU~Psc~b, a T3.5 co-moving planetary-mass companion (9--13\,\MJ) to a young (\env150\,Myr) M3 member of the AB Doradus Moving Group. The large separation between GU~Psc~b and its host star (42\arcsec) provides a rare opportunity to study the photometric variability of a planetary-mass companion. The study presented here is based on observations obtained from 2013 to 2014 over three nights with durations of 5--6\,hr each with the WIRCam imager at Canada-France-Hawaii Telescope.  Photometric variability with a peak-to-peak amplitude of $4\pm1$\% at a timescale of \env6\,hr was marginally detected on 2014 October 11. No high-significance variability was detected on 2013 December 22 and 2014 October 10. The amplitude and timescale of the variability seen here, as well as its evolving nature, is comparable to what was observed for a variety of field T dwarfs and suggests that mechanisms invoked to explain brown dwarf variability may be applicable to low-gravity objects such as GU~Psc~b. Rotation-induced photometric variability due to the formation and dissipation of atmospheric features such as clouds is a plausible hypothesis for the tentative variation detected here. Additional photometric measurements, particularly on longer timescales, will be required to confirm and characterize the variability of GU~Psc~b, determine its periodicity and to potentially measure its rotation period. 
\end{abstract}

\section{Introduction}
The study of photometric variability is a unique and powerful technique to explore the nature and dynamics of exoplanet and brown dwarf atmospheres. Notably, photometric variability is a common method to constrain the presence and evolution of clouds on an unresolved body, which can play a crucial role in shaping the observed atmospheric spectra.

In the brown dwarf regime, the detection of photometric variability is common. It has been previously demonstrated in large-sample surveys \citep{Girardin2013,Radigan2014b, Radigan2014a, Wilson2014,Metchev2015} that, as suggested by atmosphere models \citep{ShowmanKaspi2013}, a significant fraction of field brown dwarfs display large-amplitude photometric variations in the infrared, especially at the L/T transition. Notable examples include SIMP~J013656.57+093347.3 (SIMP~0136 hereafter), a T2.5 isolated object\footnote{This object, initially thought to be a field brown dwarf, was recently shown to be a likely member of the \env200 Myr old Carina-Near association, and is thus in all likelihood below the planetary-mass threshold of \env 13\,\MJ \citep{Gagne2017}.} that has been shown to display a $J$-band variation up to 6\% peak-to-peak over a period of 2.4\,hr \citep{Artigau2009,Apai2013,Metchev2013,Croll2016}, 2MASS J21392676+0220226 (2MASS~J2139 hereafter), a T1.5 displaying a peak-to-peak variability as large as 26\% in $J$ band over 7.7\,hr \citep{Radigan2012}; and WISE~J104915.57-531906.1 (Luhman~16B), a T0.5 that shows a $>10\%$ peak-to-peak amplitude variability with a \env5\,hr period in the near-infrared \citep{Gillon2013, Burgasser2014, Buenzli2015}. The cooler T6.5 2MASS~J22282889−431026 (2M~2228 hereafter; \citealp{Buenzli2012}) shows rapid variability (period 1.4\,hr) in the near- and mid-infrared bands with peak-to-peak amplitudes ranging from 1.45\% to 5.3\%. Recently, the Y0 dwarf WISE~J140518.39+553421.3 was found to be variable at the 7\% level (peak-to-peak) on a 8.5\,hr period, in the \textit{Spitzer} Infrared Array Camera (IRAC) [3.6] and [4.5] bands \citep{Cushing2016}. 

The most common explanation for the observed short-term variability of brown dwarfs is the presence of a non-uniform cloud cover in the atmosphere \citep{Apai2013}. Doppler imaging allowed us to obtain a two-dimensional map of Luhman~16B, the nearest known brown dwarf (\env2\,pc). It exhibits large-scale bright and dark regions that evolve with time and that naturally explain the observed photometric variability \citep{Crossfield2014}. Fluctuations in the temperature of the atmosphere could provide an alternative explanation for objects outside of the L/T transition \citep{Robinson2014}. Variability of 2M~2228 could be explained by a combination of patchy sulfide clouds and hot spots \citep{Morley2014, Robinson2014}. Regardless of the underlying physical mechanism, variability is in all cases primarily produced by modulation due to rotation, which brings regions with different physical properties in and out of sight. 

Variability in T dwarfs was also detected on longer timescales of days to months for most isolated objects studied over such long periods. SIMP 0136 is an extreme example with a peak-to-peak amplitude varying between less than 1\% to more than 6\% over the 6 years it was studied \citep{Artigau2009,Apai2013,Metchev2013,Croll2016}.
The evolution of variability is thought to be due to large-scale evolution of weather patterns on the surface. In our solar system, simultaneous disk-integrated and resolved photometric studies of the ice giant Neptune suggested that the short- and long-term evolution of cloud structures on the surface of planets generate variations in the photometric light curves on timescales of hours to months \citep{Simon2016}. 

Future instruments on ground-based 30 m class telescopes and JWST will allow more in-depth photometric variability studies of directly imaged exoplanets \citep{Kostov2013}. However, variability studies are currently very challenging to perform  on the majority of known giant exoplanets because of the proximity to their host star (e.g., see observations of HR~8799 by \citealp{Apai2016}). The detection of a photometric modulation in 2MASS~J12073346-3932539~b (hereafter 2M~1207~b; \citealp{Zhou2016}) illustrates that a reliable detection of the rotation-induced modulation can, however, be obtained, at least for the most favorable geometries. 

The discovery of free-floating planetary-mass objects allows extending photometric variability studies into the low-gravity regime. In their \textit{Spitzer} program studying 44 L3--T8 brown dwarfs, \citet{Metchev2015} identified a tentative correlation between low-gravity objects and large-amplitude variability. \citet{Biller2015} found the first evidence for variability for a low-gravity object, PSO~J318.5-22, a late-L planetary-mass object and member of the very young $\beta$~Pictoris moving group (\env 20\,Myr). They found the planetary-mass object to be variable with a large amplitude of 7\%--10\% peak-to peak in $J_{S}$ band, at two different epochs. Another low-gravity dwarf, the L6 WISEP J004701.06+680352.1 (W0047 hereafter), was also found to be variable with a very large amplitude in the near-infrared (8\% peak-to-peak; \citealt{Lew2016}). These low-gravity L dwarfs display larger amplitude variability than most variable field L-type brown dwarfs, suggesting that in agreement with \citet{Metchev2015}, young, dusty L planetary-mass objects could be more variable than their older counterparts of similar colors. The recent finding that the highly variable SIMP~0136 is in all likelihood also young (given its probable membership to the 200\,Myr association Carina-Near; \citealt{Gagne2017}), could suggest that this hypothesis extends for early T dwarfs. This thus calls for further observations of these young objects.

GU~Psc~b is a T3.5\plmo1 planetary-mass companion at a separation of 42\arcsec\ (2000\,au at 48\,pc) from the young M3 star GU Psc, a member of the young (\env150\,Myr; \citealp{Bell2015}) AB Doradus moving group (ABDMG). This very wide companion was identified from its distinctively red $i'-z'$ $>$ 3.5 color from the PSYM-WIDE survey carried out on Gemini-South/GMOS \citep{Naud2017a} and confirmed to be co-moving with multi-epoch WIRCam $J$-band astrometry \citep{Naud2014}. Given its estimated \Teff \env 1050\,K and the young age inferred from its membership to ABDMG, its estimated mass is at the high-end of the planetary-mass regime (9--13\,\MJ). 

GU~Psc~b has a similar mass to closer-in giant exoplanets revealed by high-contrast imaging and isolated planetary-mass objects. Besides, it shares similar spectral features to much older and massive field brown dwarfs at the L/T transition. Its study allows us to investigate for connections between these two types of objects.
As it is one of a few dozen exoplanets that have been directly imaged and one rare case among those that were detected without the aid of adaptive optics, GU~Psc~b presents an opportunity to study the light curve of an exoplanet similarly to what was done for older early T dwarfs with current instruments. 

Previous observations presented by \citet{Naud2014} showed no $J$-band variability above 150\,mmag (at a 3$\sigma$ confidence level) over 3 epochs spanning 11 months. 
This paper presents the first dedicated monitoring of the broadband photometric variability of GU~Psc~b. In \S \ref{sec:obsred}, observations obtained at the Canada-France-Hawaii Telescope (CFHT) in 2013--2014 are described. The light curves obtained are presented in \S \ref{sec:res}. In \S \ref{sec:anal}, the analysis of these light curves is detailed. Finally, the importance of this result in light of other recent variability studies is discussed in \S \ref{sec:analdis}, and future observations that could reveal additional insights on the atmospheric dynamics of planetary-mass objects are suggested.

\section{Observations and Data Reduction}
\label{sec:obsred}

Observations of GU~Psc~b were obtained through Director's discretionary time at the CFHT\footnote{Run IDs 13BD91 and 14BD88.} with the near-infrared camera WIRCam \citep{Puget2004}. Since rotation periods for young low-mass companions are still largely unknown, the longest continuous observation span that could be secured continuously on CFHT was requested. Three 5--6\,hr $J$-band observing periods on three different nights were obtained: on 2013 December 22, and on 2014 October 10 and 2014 October 11 (see Table \ref{tab:obslog}). Long exposure times (50\,s in 2013 December and 60\,s in 2014 October) were used in order to achieve the best possible signal-to-noise for this faint target ($J_{MKO}=18.12$, \citealp{Naud2014}). 

WIRCam is equipped with four 2048$\times$2048 pixels HAWAII 2RG detectors (pixel scale of 0\farcs307/pixel) spanning a field of view of 20\arcmin$\times$20\arcmin. The target was kept approximately at the same position in an area clean of cosmetic defects in the northwest detector for the complete duration of a given observation epoch, using the WIRCam staring mode \citep{Devost2010}. WIRCam was purposely slightly defocused relative to the best focus position of the primary mirror, just enough to stabilize the PSF in the event of changing seeing conditions while keeping the PSF Gaussian and ensuring that the flux remained significantly above the sky level. The dialed defocus was 0.20\,mm and never drifted more than 0.05\,mm from the telescope model position, which ensured that the PSF had a minimum FWHM of $2.2\pm0.5$~pixels. Appropriate master twilight flats and darks were obtained in the standard CFHT calibration sequences. In 2013 December, sky observations were obtained but they did not allow us to improve the quality of the results and were thus not used. No further sky observations were obtained in 2014 October. 

The IDL Interpretor of WIRCam Images (\texttt{I'iwi v2.1.200}\footnote{See \url{http://www.cfht.hawaii.edu/Instruments/Imaging/WIRCam/IiwiVersion1Doc.html}}) was used for preprocessing of the raw data including dark subtraction, flat fielding with twilight flat, bad pixel mapping and nonlinearity correction.

\begin{table}[htbp]
 \newcolumntype{A}[1]{>{\arraybackslash}m{#1}} 
\caption{Observation Log}
\label{tab:obslog}
\begin{center}
\begin{tabular}{cllc}
\hline \hline      
 Date & Time Start & $t_{\rm{exp}}$ & Total Duration \\
 (UTC) & & (s)& (hr) \\
\hline
2013 Dec 22 &04:41:22& 50.0&5 \\
2014 Oct 10 & 8:40:37 & 60.0&6 \\ 
2014 Oct 11 & 7:07:20 &60.0& 6 \\
\hline
\end{tabular}
\end{center}
\end{table}

\section{Results}
\label{sec:res}

\subsection{Raw Light Curves}
\begin{figure}[htbp]
\begin{center}
\includegraphics[width=8.5cm]{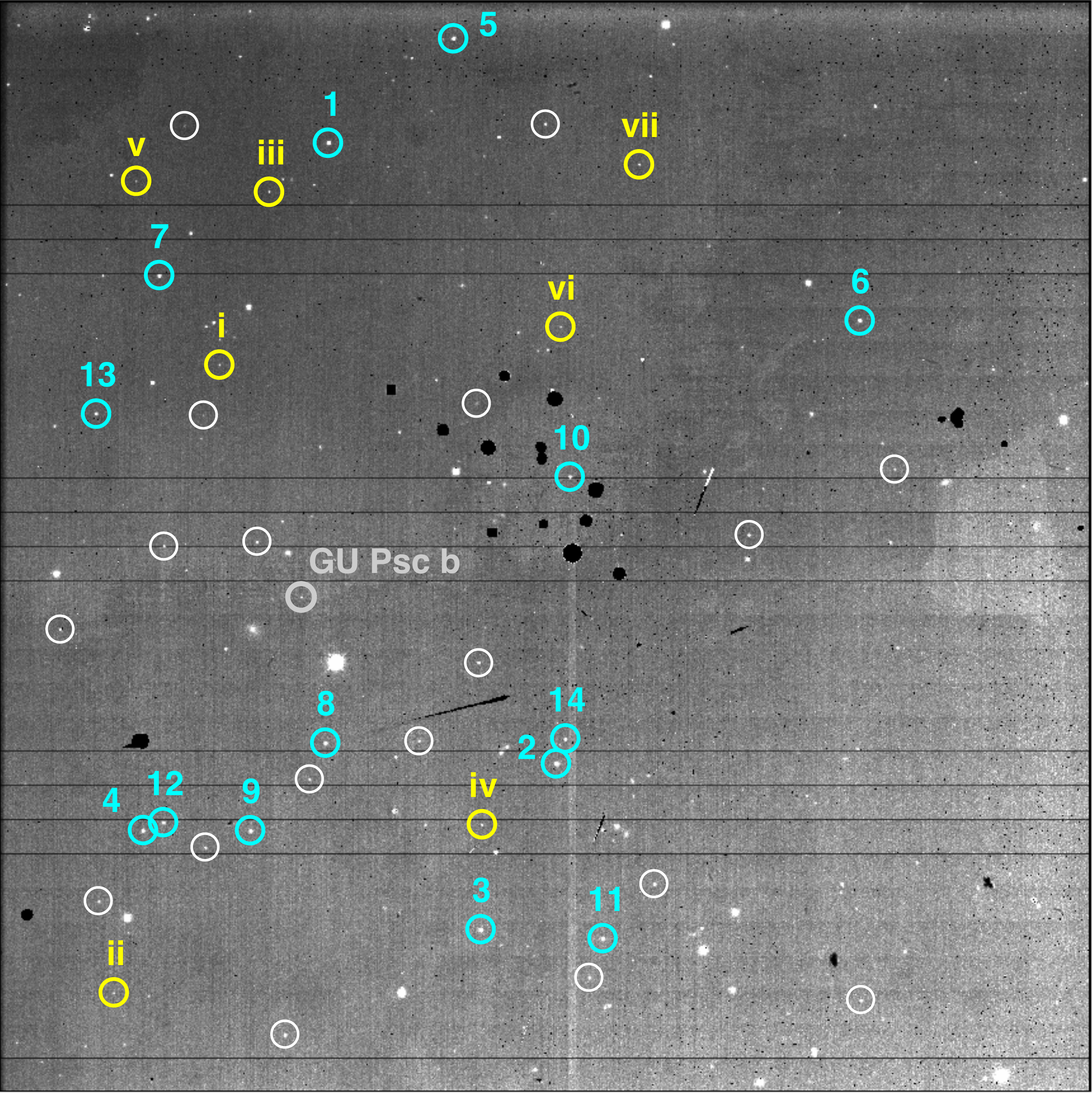}
\caption{Deep image of the field observed in $J$ band with WIRCam (stack of the 6\,hr 2014 October 11 epoch). The field shown is the northwest detector of WIRCam ($\#77$) and is 10\arcmin\ on a side. GU~Psc~b and the 40 stars considered in the analysis are circled. Among these stars were selected 14 reference stars, high SNR stars that are in the 2MASS catalog, which were used to correct the light curves. They are the same for the three observation runs. They are identified in cyan and numbered 1--14. Seven comparison stars, with a brightness similar to GU~Psc~b, used to validate the results, are circled in yellow and numbered \textrm{i} to \textrm{vii}.} 
\label{fig:findingchart}
\end{center}
\end{figure}

IDL procedures were used to perform aperture photometry on the target for each individual exposure. The same operation was performed on 40 stars identified in the field, located close to the target on the same detector and with a brightness between 0.1 and 10 times that of the target (see Figure \ref{fig:findingchart}). 

The position and FWHM of all stars at each time step were first determined using the IDL procedure \texttt{MPFIT2DPEAK}, which adjusts a 2D Gaussian profile at the approximate position identified manually. The IDL procedure \texttt{APER} was then used to do aperture photometry and extract the raw light curves. An aperture fixed in size (rather than a multiple of the changing FWHM) located at the median position was adopted all along the observations of a given epoch. This aperture proved to generate the most stable light curves, even though the precise position of the stars and the FWHM of their PSF varied during the observation (see Figure \ref{fig:extparam} for 2011 October 11, and Figures \ref{fig:extparam2} in the Appendix for 2013 December 22 and 2014 October 10).

\begin{figure*}[htbp]
\begin{center}
\includegraphics[width=15.5cm]{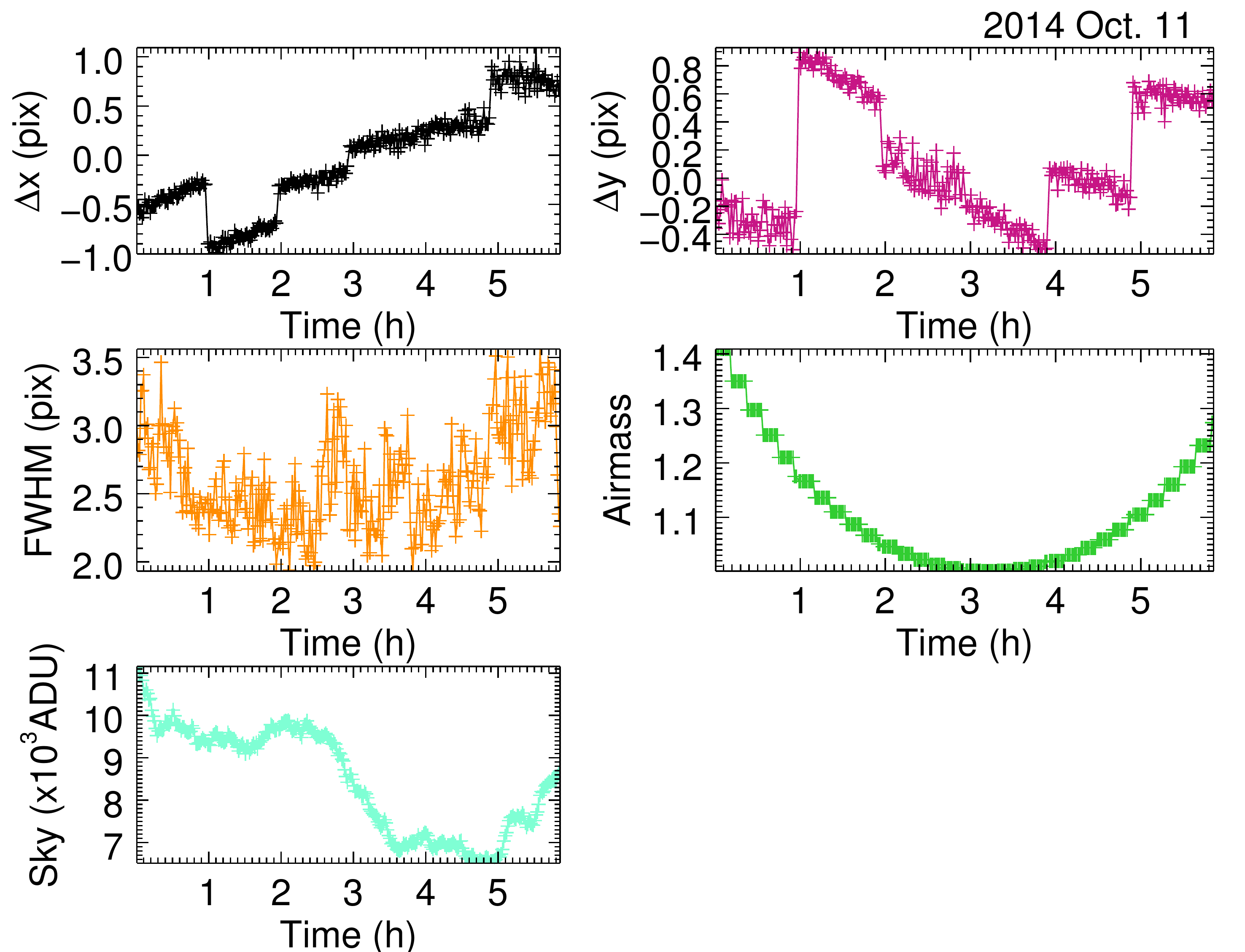}
\caption{Variation of different external parameters during the 2014 October 11 epoch. The variation of the $x$ and $y$ positions on the chip (median of all stars), FWHM, airmass, and sky level (ADU) are shown. Similar figures for the two other nights are shown in the Appendix.}
\label{fig:extparam}
\end{center}
\end{figure*}

Apertures of various sizes ranging from a radius of 1 to 8 pixels were tested. An aperture of 3 pixels was selected. This aperture is small enough to minimize the Poisson noise for our faint target (sky-background dominated) but large enough to avoid systematic errors due to the loss of flux caused by slight displacements of the star and variations in the seeing at a given epoch and across the field. Annuli with inner and outer sky radii of 4.5 and 9 pixels were used to measure the sky contribution.
The measured flux was converted to a relative flux by dividing the entire light curve of every star by its median. The raw light curve generated for GU~Psc~b on 2014 October 11 is shown in Figure \ref{fig:raw}. The raw curves for the two other nights are shown in Figure \ref{fig:raw2} of the Appendix. Among the 40 stars initially identified for which a raw light curve was extracted (all stars circled in Figure \ref{fig:findingchart}), seven stars with a brightness similar to that of GU~Psc~b (85\%--130\% of the flux of the target, identified by yellow circles and roman numerals in Figure \ref{fig:findingchart}) were selected as comparison stars. Their median raw light curve is also displayed in Figure \ref{fig:raw}. A set of 14 reference stars were also selected, consisting of bright stars that have a high signal-to-noise ratio (larger than \env 60 per measurement), that are not obviously variable and that are listed in the 2MASS catalog (identified by cyan circles and arabic numerals in Figure \ref{fig:findingchart}). Their characteristics are listed in Table \ref{tab:refstar} and their median raw light curve is shown in Figure \ref{fig:raw}. The reference stars are used to correct the raw light curves of GU~Psc~b and comparison stars (see section \ref{sec:anal} for more detail on the procedure used to do so).

\begin{figure*}[htbp]
\begin{center}
\includegraphics[width=15cm]{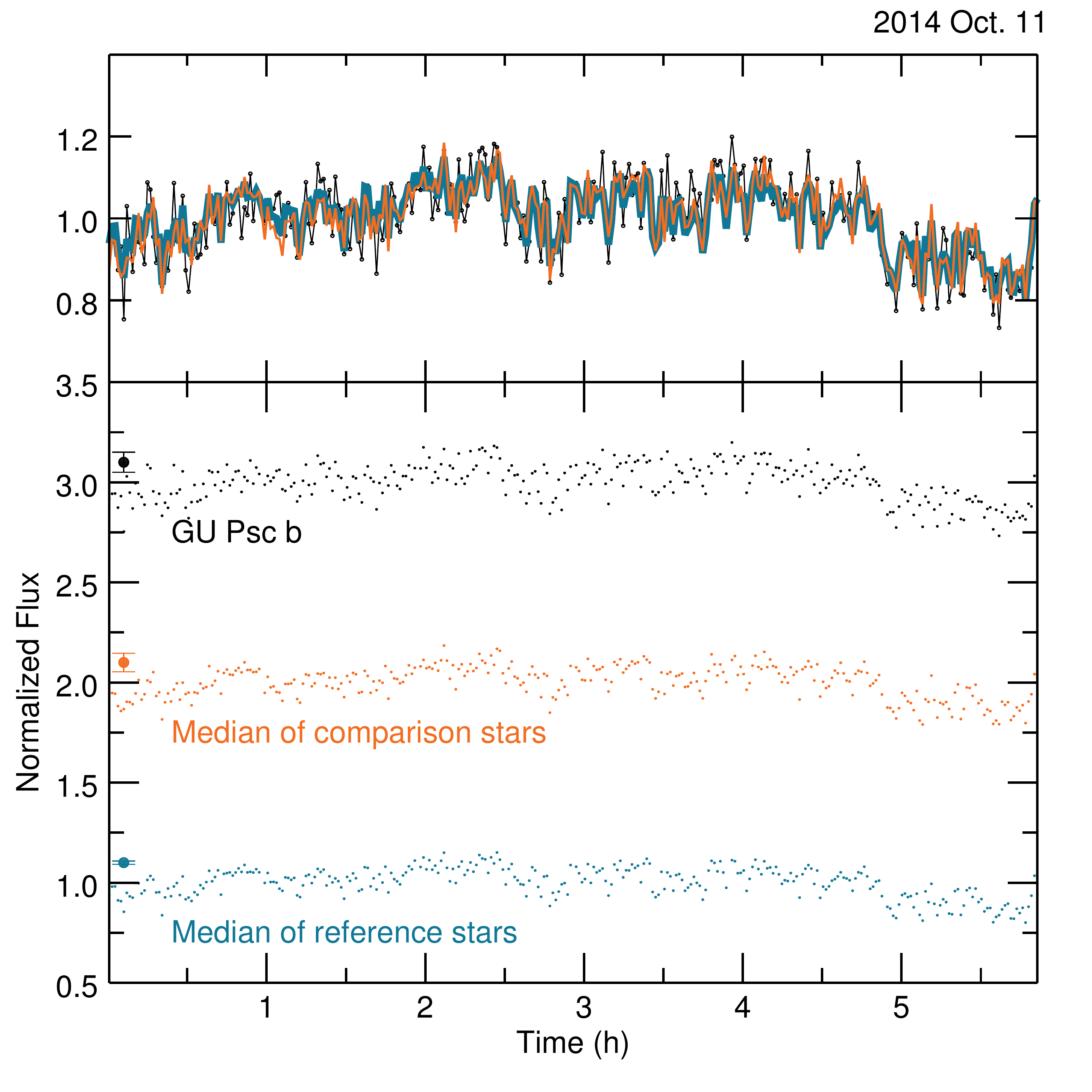}
\caption{Bottom panel: raw normalized light curves for GU~Psc~b (top curve, in black), the median of comparison stars (middle curve, in orange) and the median of all reference stars (bottom curve, in cyan), for 2014 October 11. GU~Psc~b and comparison star median curves have been offset for clarity. Top panel: the same curves, overplotted over one another, GU~Psc~b is the black curve, with dot symbols, the median of the reference stars is the thick dark cyan line, and the median of the comparison stars is the thin orange pale line. The raw curves for the two other nights are shown in the Appendix.}
\label{fig:raw}
\end{center}
\end{figure*}

\begin{table*}[htbp]
 \newcolumntype{A}[1]{>{\arraybackslash}m{#1}} 
\caption{Selected Reference Stars}
\label{tab:refstar}
\begin{center}
\begin{tabular}{ccccccc}
\hline \hline      
 2MASS & ID\tnma & Distance & R.A. & Decl. & \multicolumn{2}{c}{2MASS} \\
 \cline{6-7}
 Designation &  & from Target (\arcmin) & (deg) & (deg) & $J$  & $K_{s}$\\
\hline
     J01123542+1708511  & 1  & 4.33 & 18.148    & 17.148    &  15.514  & 14.5490\\      
     J01122636+1702556	 & 2 &  2.90 & 18.110    &  17.049   &  15.481  &  14.8010\\
     J01122937+1701205  & 3 &  3.61 & 18.122    &  17.022   &  15.638   &14.8110 \\
     J01124286+1702189  & 4 & 2.69 & 18.179    &  17.039   &  15.808  &   15.4340\\
     J01123039+1709489	 & 5 &  5.48 & 18.127    &  17.164   &  15.852  &  15.2760 \\
     J01121425+1707070  & 6 &  5.91 & 18.059    &  17.119   &  15.825  &  15.1590\\
     J01124214+1707359  & 7 &  3.35 & 18.176    &  17.127   &  16.133  &  15.4240\\
     J01123560+1703082  & 8 &  1.41 & 18.148    &  17.052   &  16.099  &  15.2590\\ 
     J01123859+1702178  & 9 &  2.29 & 18.161    &  17.038   &  16.027  &  15.3530\\ 
     J01122580+1705389  & 10 & 2.79 & 18.108    &  17.094   &  16.408  &  15.8430\\
     J01122456+1701154	& 11  & 4.34 & 18.102    &  17.021   &  16.520  & 16.1250 \\
     J01124208+1702228 & 12  & 2.53 & 18.175    &  17.040   &  16.921  &  15.4570 \\
     J01124468+1706169 & 13  & 2.63 & 18.186    &  17.105   &  16.662  &  15.5090 \\
     J01122599+1703093	& 14  & 2.86  &  18.108   &   17.053  &   16.333 &  15.4090\\
\hline
\multicolumn{5}{A{8.3cm}} {\textbf{Note} \tnma The ID refers to Figure \ref{fig:findingchart}}\\ 
\end{tabular}
\end{center}
\end{table*}

The variations of several external parameters such as the position on the detector, seeing, airmass and temperature were also monitored to study their possible effect on the data (Figure \ref{fig:extparam}). The variation of positions, which were found to be similar for stars on the same detector (Figure \ref{fig:extparam} shows the median of all stars), arises partly from the WIRCam science acquisition sequence. At the end of each observing sequence (about 1\,hr), the position of the pointing was adjusted. That can be seen, for example on 2014 October 11 by the variation of the $\rm{y}$ position every hour.  The measured FWHM of the PSF also varied during a given epoch due to the changing seeing. The median FWHM of the PSF is 2.4 pixels for the first night, 2.3 pixels for the second night, and 2.5 pixels for the third. On the first night, an important degradation of the seeing can be seen \env4.5\,hr after the beginning of the observations. Observations were obtained at airmasses below 2.0.  

\section{Analysis}
\label{sec:anal}
\subsection{Principal Component Analysis}
A common procedure to eliminate instrumental noise common to all observed stars is to divide the raw light curves by a reference curve, which is usually built from the mean or median of carefully selected reference stars \citep{Radigan2014a}. However, in the present case, such a procedure leaves residual variability that is not likely of astrophysical nature, as patterns can be recognized in the ``corrected'' light curves of both GU~Psc~b and comparison stars, and seem correlated with external parameters. 

A principal component analysis (PCA; \citealp{Jolliffe2002}) was used to efficiently eliminate this common instrumental noise. The raw light curves of the 14 reference stars were used as inputs in the IDL function \texttt{PCA}, which computes their covariance matrix and finds its eigenvectors (or Principal Components; PC) and eigenvalues. Principal components, which are orthogonal by construction, are ordered in decreasing contributions to the variance of the sample set. A few PCs usually contribute to most of the variance in the data, as was observed here, hence those with less significance were ignored. A Scree plot was used to determine how many PC to retain. This plot displays eigenvalues in decreasing order, and only the PCs that have a value in the steep decline, before a plateau is reached, are kept. For the three nights, 3/14 PCs were used, which accounted for more than 98\% of the variance in all three cases. These PCs are displayed in Figure \ref{fig:PC} for the third epoch, and in the Appendix (Figure \ref{fig:PC2}) for the two other. It was verified that retaining one more or one less PC does not affect the results significantly. 

\begin{figure*}[htbp]
\begin{center}
 \includegraphics[width=18.cm]{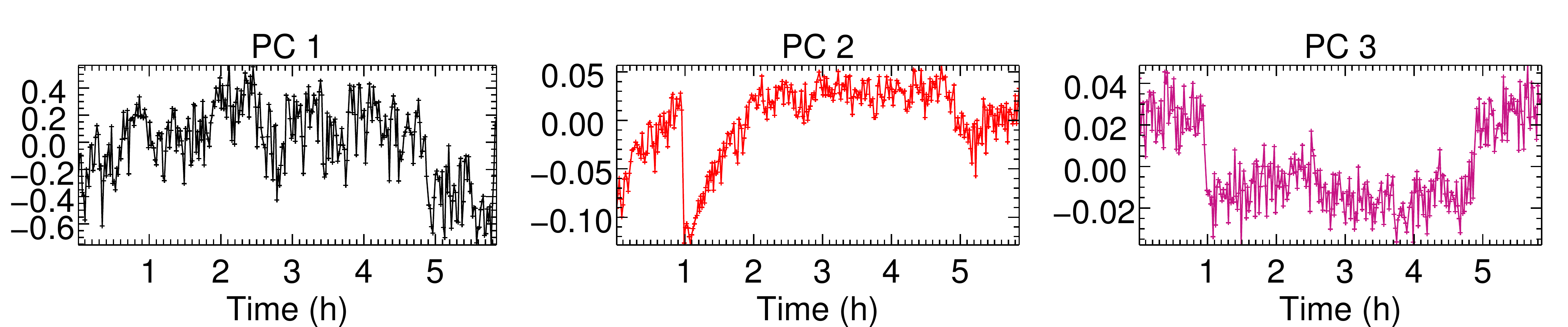}
\caption{Most important principal components for 2014 October 11, obtained from the light curves of the 14 reference stars (identified with arabic numbers in Figure \ref{fig:findingchart}). Similar figures for the two other nights are shown in the Appendix.}
\label{fig:PC}
\end{center}
\end{figure*}

Comparing the principal components to the variation of external parameters (shown in Figure \ref{fig:extparam}) can shed light on what has the most effect on the observed common variability. The strongest effect is the variation of the seeing. The first PC shows an anti-correlation of $>$95\% with the measured FWHM for the three epochs. A few other external parameters have recognizable patterns that can be found in the principal components. The jumps that are seen approximately at 1 and 5\,hr in the $\Delta \rm{y}$ curve of Figure \ref{fig:extparam} (2014 October 11) can be seen in the second PC. Other varying parameters could have an impact as well, but the effect of the smoothly varying ones (sky level, airmass) are harder to disentangle.

\subsection{Autocorrelation Analysis}
\begin{figure}[htbp]
\begin{center}
\includegraphics[width=8.5cm]{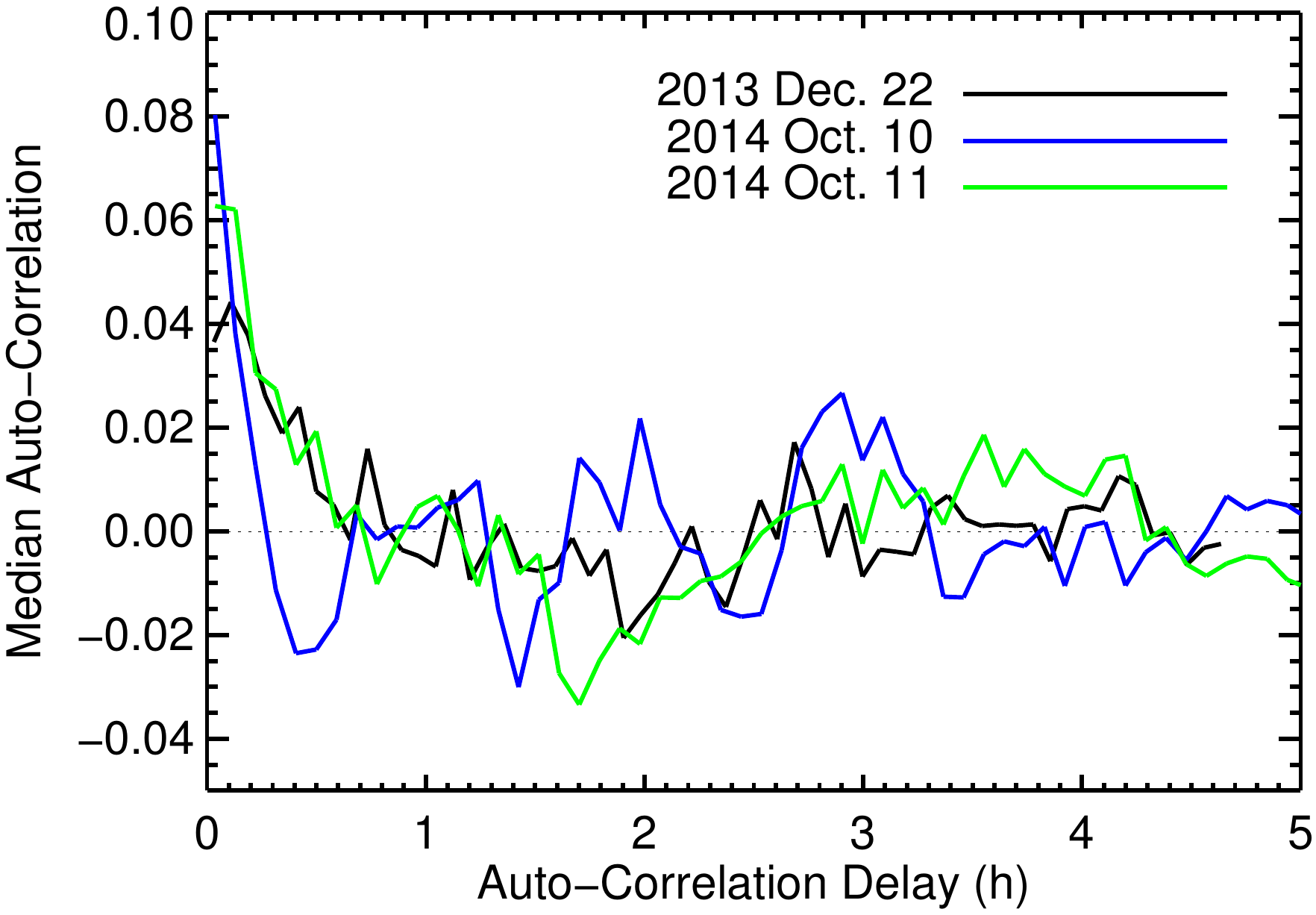} 
\caption{Result of the autocorrelation analysis that was carried out. The reference star light-curve residuals are used, after the removal of the instrumental noise with PCA analysis. The median of the autocorrelation curves of the 14 reference stars is shown for each epoch.}
\label{fig:ac}
\end{center}
\end{figure}

The following procedure was applied to the reference stars to evaluate the presence of correlated noise in the light curves. An optimal reference light curve was built for each of the 14 reference stars using the linear combination of the three first principal components, which minimizes the RMS of residuals in the corrected light curve.
In each case, the reference star in question was excluded from the determination of the PC. Corrected light curves were obtained by subtracting the optimal reference light curve from the raw light curves. An autocorrelation analysis was carried out on the residuals of these corrected light curves. Figure \ref{fig:ac} shows the autocorrelation curves at all epochs for the median of the 14 stars. The analysis showed no evidence for noise correlation on timescales longer than 30 minutes. For timescales significantly longer than 30 minutes and variability amplitudes larger than the RMS of bright stars (0.5\% peak-to-peak), the noise can therefore be assumed to be white.

\subsection{Correcting for Instrumental Noise and Variability Modeling}
\label{subsec:model}
\begin{figure*}[htbp]
\begin{center}
\includegraphics[width=13.5cm]{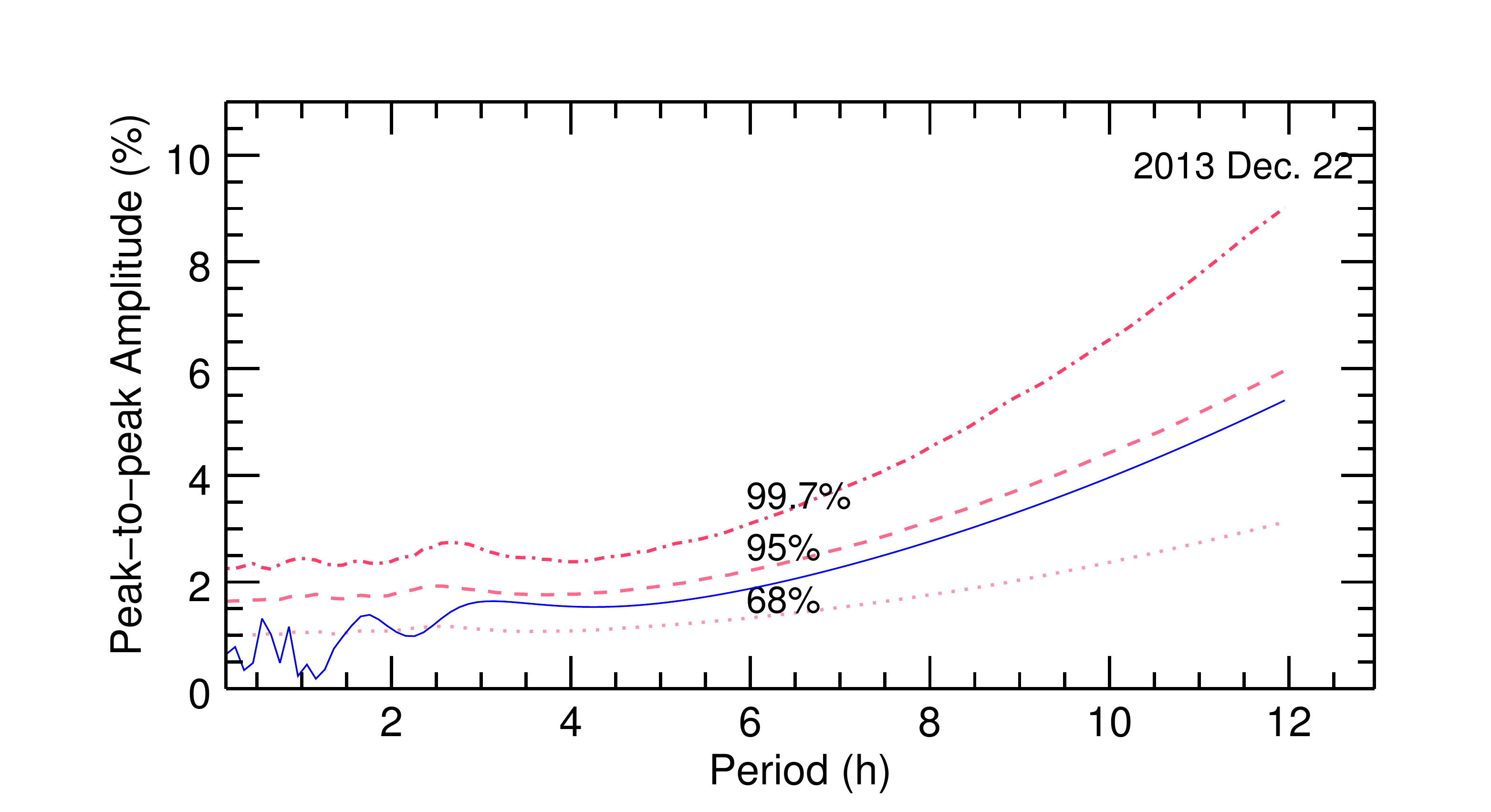}

\includegraphics[width=13.5cm]{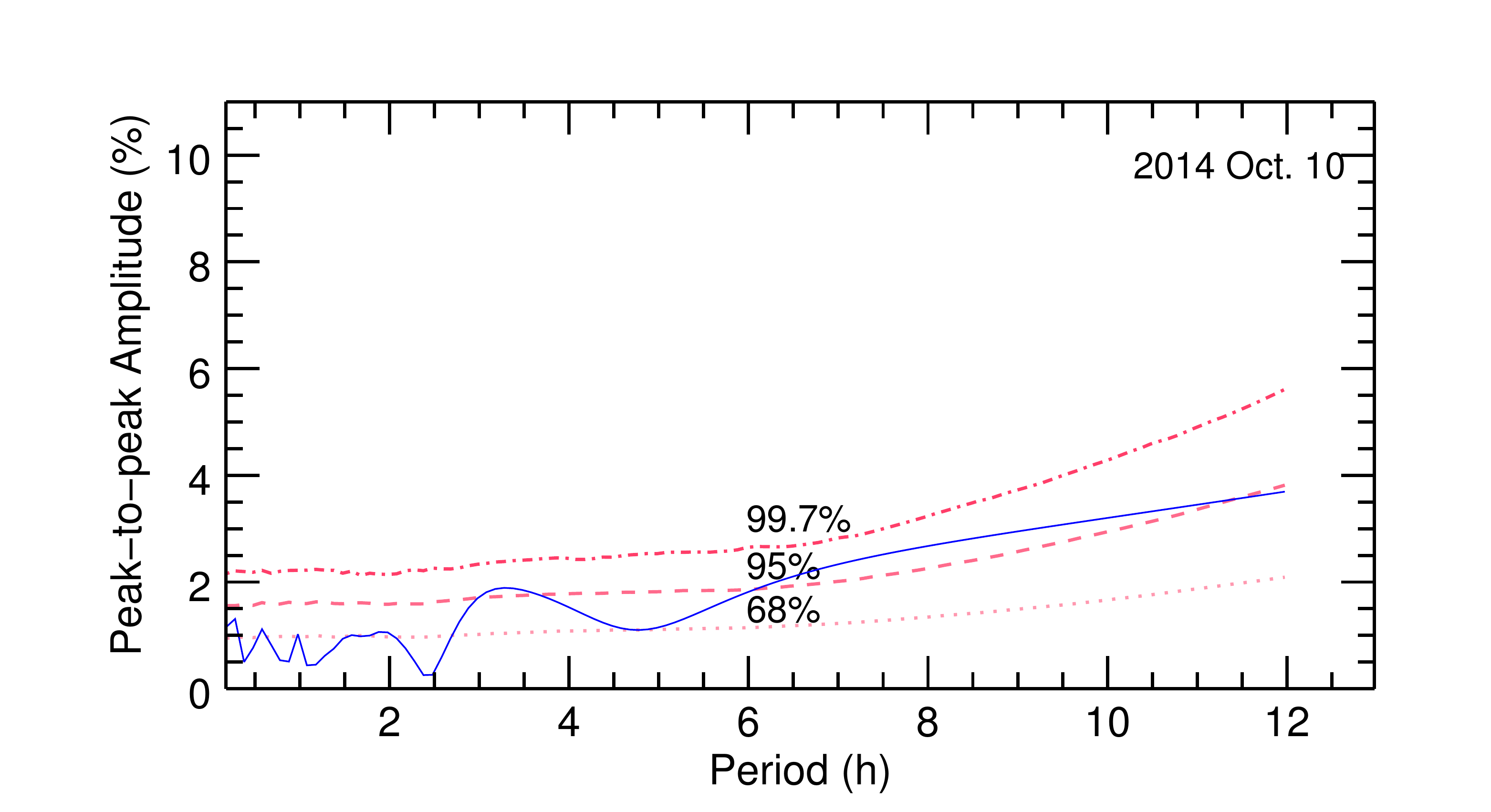}

\includegraphics[width=13.5cm]{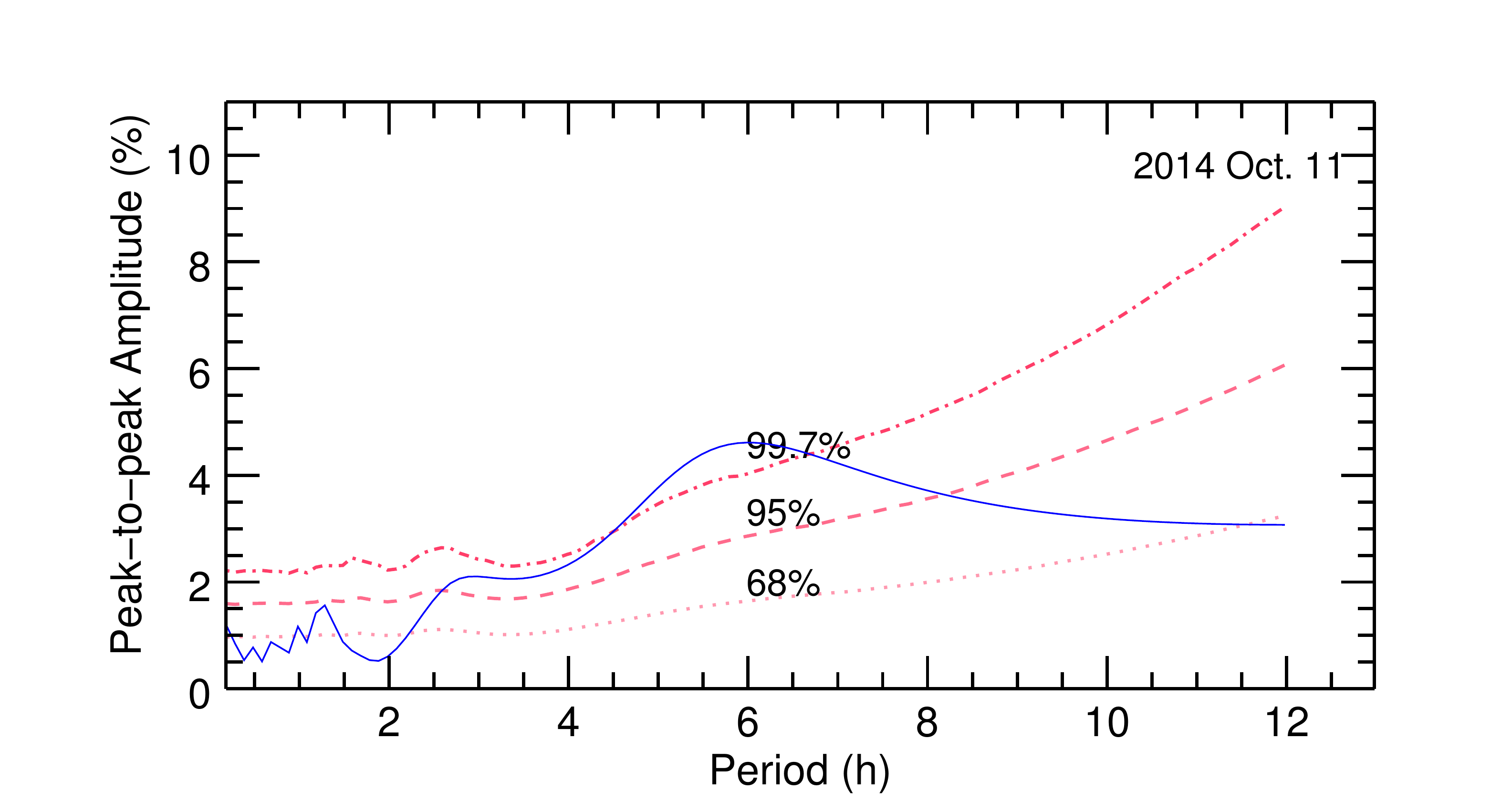}
\caption{Amplitude of the best-fit model for all periods tested on the light curve of GU~Psc~b (solid line), for the three epochs. We use the unbinned data in the fit and simulation. The 68\%, 95\%, and 99.7\% detection limits are also shown (dotted, dashed, dash-dotted lines). No significant variability is detected (above the 99.7\% limit) for the first and second nights. On the third night, a 4--5\% peak-to-peak amplitude is seen around 6\,hr, just above the 99.7\% limit. }
\label{fig:periodogram}
\end{center}
\end{figure*}

Instrumental noise and astrophysical variability can often be correlated. A joint model was used to simultaneously find the scale of the PCs (to remove instrumental noise) and find the best fit to a possible intrinsic variability. This procedure allows us to avoid problems caused by a correlation between the two. The model in question is described by
\begin{equation}\label{eq:model}
m_i = A\ \cos\left[ \frac{2 \pi}{P} (t_i+ t_0) \right]\ +\ \sum_{k=1}^{3} B_k\ \omega_{ik} + C,
\end{equation}
where $A$ is the amplitude of intrinsic variability, $P$ is the period, and $t_0$ is the time offset. The three principal components (noted $\omega_{ik}$) are scaled by $B$. 
Even if the light curves of many variable brown dwarfs and planetary-mass companions are not sinusoidal, the present model was selected as a simplistic choice to search for stellar variability without overfitting the data. The model is rewritten in the following linear form to allow for a matrix inversion:
\begin{equation}\label{eq:model2}
m_i = A_{C}\ \cos\left( \frac{2 \pi}{P} t_i \right)\ + A_{S}\ \sin\left( \frac{2 \pi}{P} t_i \right)\ +\ \sum_{k=1}^{3} B_k\ \omega_{ik} + C,
\end{equation}
with $A=\sqrt{A_{S}^{2}+A_{C}^{2}}$ and 
\begin{equation}
t_0=\frac{P}{2\pi}\arccos \left( \frac{A_{C}}{ \sqrt{A_{C}^{2}+A_{S}^{2}}} \right).
\end{equation}
The optimal parameters $A$, $P$, $t_0$, $B_{1}$, $B_{2}$, $B_{3}$ and $C$ are determined for periods ranging from \env 10\,minutes up to twice the total duration of the observation period (12\,hr), in step of 6\,minutes. The resulting peak-to-peak amplitude ($2A$) for each trial period are shown in Figure \ref{fig:periodogram} for GU~Psc~b (solid blue line).  

A Monte Carlo simulation was performed to evaluate the detection limits: the data points were shuffled randomly 10,000 times, and each new set of data was fitted again, yielding an amplitude for each trial. The 68\%, 95\%, and 99.7\% lines on Figure \ref{fig:periodogram} for GU~Psc~b (dotted, dashed, dash-dotted red lines) represent the amplitudes below which 68\%, 95\%, and 99.7\% of the simulations are found.

A \env4--5\% peak-to-peak amplitude signal is detected at a period of \env6\,hr from the analysis for the third epoch. No significant variability is detected for GU~Psc~b on the first two epochs. A similar analysis for the comparison stars was also carried out. Five of the seven stars show no variability at this level of significance on the three epochs. Star \textrm{v} shows a $>3 \sigma$ variation on the same date, around a period of 4.7\,hr and with a peak-to-peak amplitude of 4\%, and a \env$3 \sigma$ variation on 2014 October 10, with a period and an amplitude that are a bit smaller (3.9\,hr, 2.7\%). Star \textrm{i} shows a $>3 \sigma$ variation at the third epoch as well, with a period of 2.7\,hr and a peak-to-peak amplitude of 2.5\%.

\subsection{Monte Carlo Markov Chain Analysis}
A Monte Carlo Markov Chain (MCMC) analysis was carried out on the GU~Psc~b 2014 October 11 light curve to assess confidence in the detection of variability and explore the parameter space and possible correlations between them. The model shown in equation \ref{eq:model} was used. Since the autocorrelation analysis showed no correlated noise, an uncorrelated Gaussian noise likelihood function was adopted, given by 
\begin{equation}\label{eq:loglike}
\log \mathcal{L} = -\frac{1}{2} n \log (2\pi) - n \log \sigma - \frac{1}{2} \sum_{i=1}^{n} \left( \frac{d_i - m_i}{\sigma} \right)^2,
\end{equation}
where $n$ is the number of observations, $d_i$ the photometric measurements, and $m_i$ is the model as given by Equation \ref{eq:model}. An uncertainty $\sigma$, that is assumed constant through the entire observation period, was left as a free parameter in the model fits. The model has eight free parameters: $A$, $P$, $t_0$, $B_1$, $B_2$, $B_3$, $C$ and $\sigma$.  Uniform priors were adopted but restrictions were applied: $A$ was forced to be positive, $P$ between 0 and 12\,hr, $t_0$ between $-$10 and 20\,hr, $B_k$, and $C$ forced to be between $-$10 and 10 and $\sigma$ was forced to be positive. The MCMC routine of \citet{Rowe2014} was adopted for the analysis. It uses the Metropolis-Hastings algorithm with a hybrid Gibbs and DE-MCMC sampler to efficiently handle correlated variables as described in \citet{Gregory2011}.

\begin{figure*}[htbp]
\begin{center}
\includegraphics[width=16.5cm]{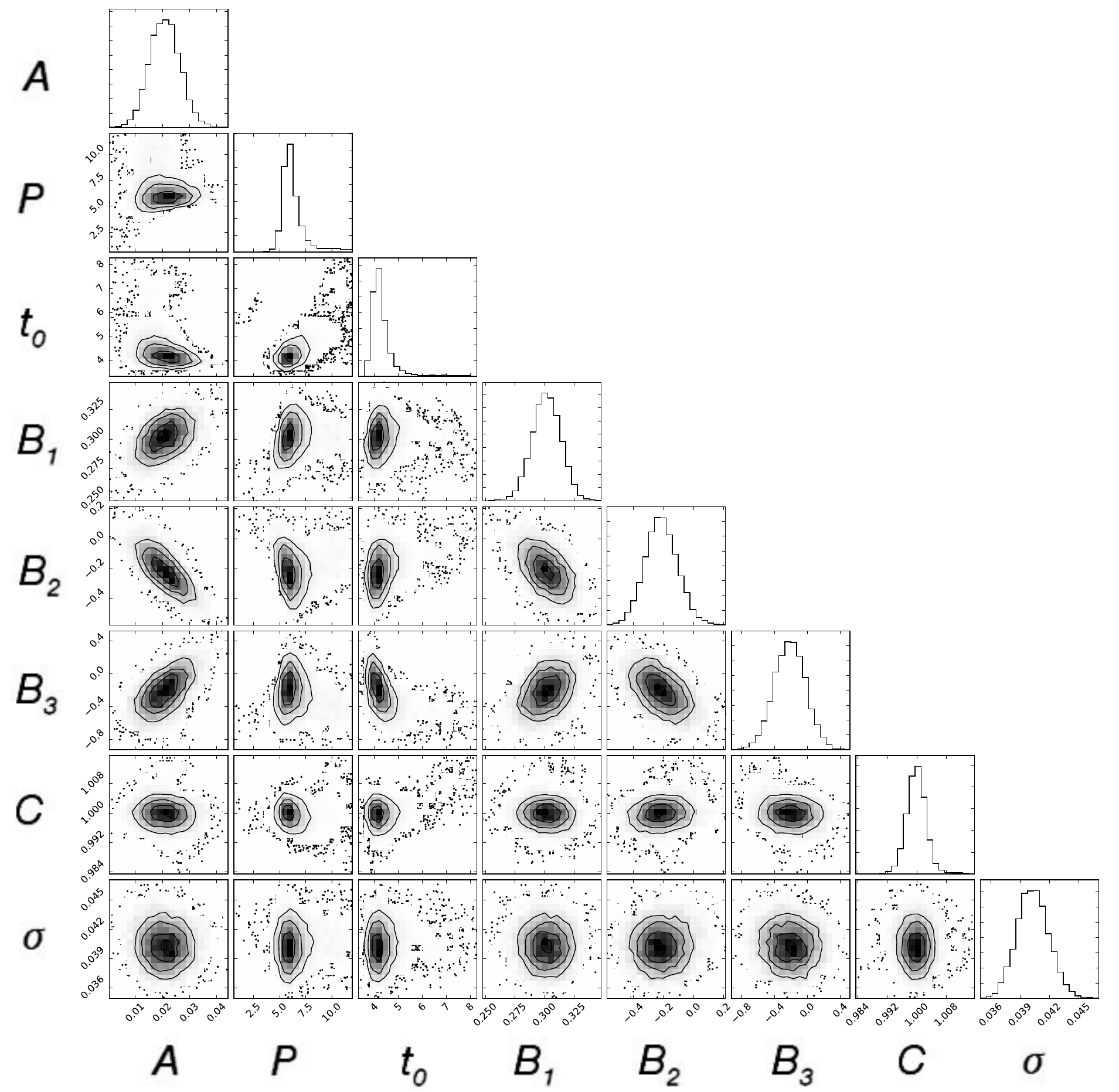}
\caption{Distributions of parameters (the amplitude $A$, the period $P$, the time offset $t_{0}$, the scale of the three principal components $B_{k}$, the constant $C$ and the constant uncertainty $\sigma$) from the MCMC analysis of GU~Psc~b light curve on 2014 October 11. Scatter plots are produced for each pair of parameters and are a good diagnostic to reveal correlations between model parameters. For example, the amplitude $A$ is correlated with $B_2$. Histograms of the parameter distributions are also shown. They provide for each parameter a representation of the posterior distribution. }
\label{fig:triplot}
\end{center}
\end{figure*}

Three chains (sequences of  ''states'' that have a given value for each parameter) with lengths of 50,000 were generated.  A visual examination of the chains showed good mixture and the Gelman-Rubin convergence criteria yielded $R_c=1.01$ or lower for all fitted parameters (a $R_{c}<1.1$ is a good indicator that convergence was reached; \citealp{Gelman1992, Brooks1998}). Figure \ref{fig:triplot} displays histograms of the chain values for each parameter and scatter plots to unveil potential correlations between various parameters for GU~Psc~b on 2014 October 11.  While it is apparent from Figure \ref{fig:triplot} that some parameters show some degree of correlation (e.g. $B_2$ and $A$), overall each parameter is characterized by a well-defined, peaked distribution. This shows that parameters are well bound given the light-curve model adopted in our analysis. From this analysis, the variability of GU~Psc~b, as observed at this epoch, can be described by a cosine with a peak-to-peak amplitude of $4\pm1$\% at a timescale of \env 6\,hr. The MCMC calculations show that the posterior distribution for the period has a long tail toward large values, with larger periods becoming increasingly unlikely. At periods much longer than the span of observations, the cosine becomes increasingly degenerate with a straight line. The upper limit was increased from 12 to 20\,hr and that did not change the 1$\sigma$ uncertainties. However, it does demonstrate that it is not possible from the current observations to determine the true periodicity of the variability of GU~Psc~b. The mode and 1$\sigma$ (68.27\%) confidence intervals for all model parameters are reported in Table \ref{tab:modelpars}. This result is compatible to what was obtained in subsection \ref{subsec:model}.  

\begin{table}[htbp]
\newcolumntype{A}[1]{>{\arraybackslash}m{#1}} 
\caption{Model Parameters for GU~Psc~b on 2014 October 11 from MCMC analysis}
\label{tab:modelpars}
\begin{center}
\begin{tabular}{lccc}
\hline \hline      
Parameter & Median & +1$\sigma$  & -1$\sigma$ \\
\hline
$2A$ (\%) & 4 & +1 & -1  \\
$P$ (hours) & 5.9 & +0.7 & -0.7   \\
$t_0$ (hours) & 4.1 & +0.4 & -0.2  \\
$B_{1}$ & 0.30 & +0.02 & -0.01  \\
$B_{2}$ & -0.2 & +0.1 & -0.1  \\
$B_{3}$ & -0.2 & +0.2 & -0.2  \\
$C$ &1.000 & +0.002 & -0.003  \\
$\sigma$ & 0.040 & +0.002 & -0.002 \\
\hline
\end{tabular}
\end{center}
\end{table}


The MCMC routine was repeated for the two other epochs and each of the seven comparison stars for all epochs. Median values of peak-to-peak amplitudes ($2A$) ranges from 1\% to 4\%, where that of GU~Psc~b at the third epoch is the largest. 

\begin{figure*}[htbp]
\begin{center}
\includegraphics[width=15cm]{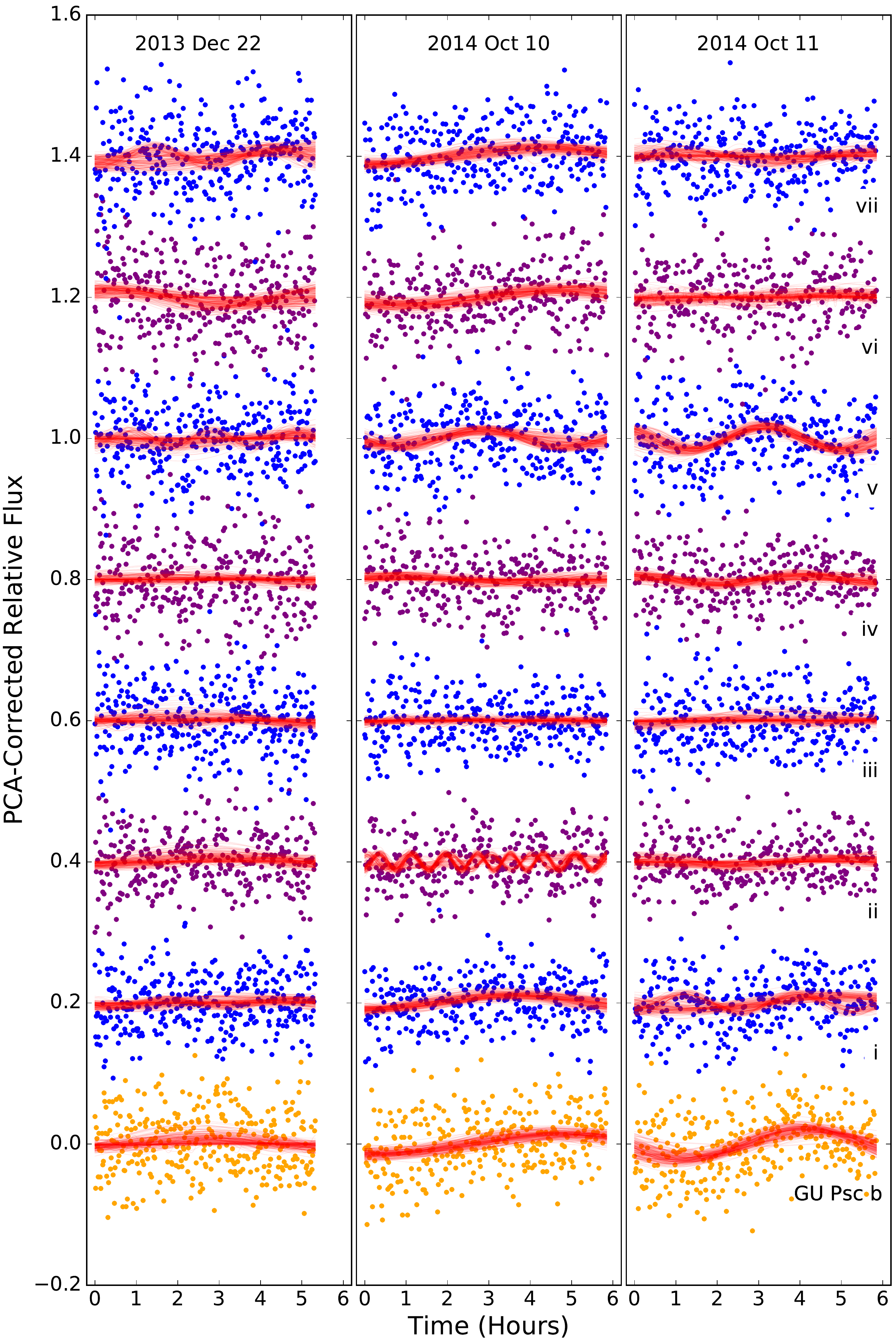}
\caption{Corrected light curve for GU~Psc~b and the 7 comparison stars, for the three epochs. Common instrumental noise was removed using PCs with coefficients $B_{k}$ based on the median MCMC values (yellow points for GU~Psc~b, blue and red points for comparison stars). Two hundred random states of the parameters of the cosine model are shown, overplotted in red, for each light curve. The \env6\,hr signal can be seen readily for this corrected light curve for GU~Psc~b on 2014 October 11.}
\label{fig:modelfit}
\end{center}
\end{figure*}

Figure \ref{fig:modelfit} shows, for GU~Psc~b and all comparison stars, the raw light curves that have been corrected using the linear combination of the PCs with the median $B_{k}$ coefficients from the MCMC analysis. Random states of the MCMC chains for the cosine model are overplotted in red. For GU~Psc~b, a significant signal can be seen by eye on the corrected light curve for 2014 October 11. A similar amplitude, albeit of longer period also seems visible on 2014 October 10, but this is not statistically significant. The light curves of most comparison stars appear less variable, by such an inspection. A notable exception is the third epoch of star \textrm{v}. 

\subsection{Bayesian Information Criterion}
\begin{figure*}[htbp]
\begin{center}
\includegraphics[width=16.5cm]{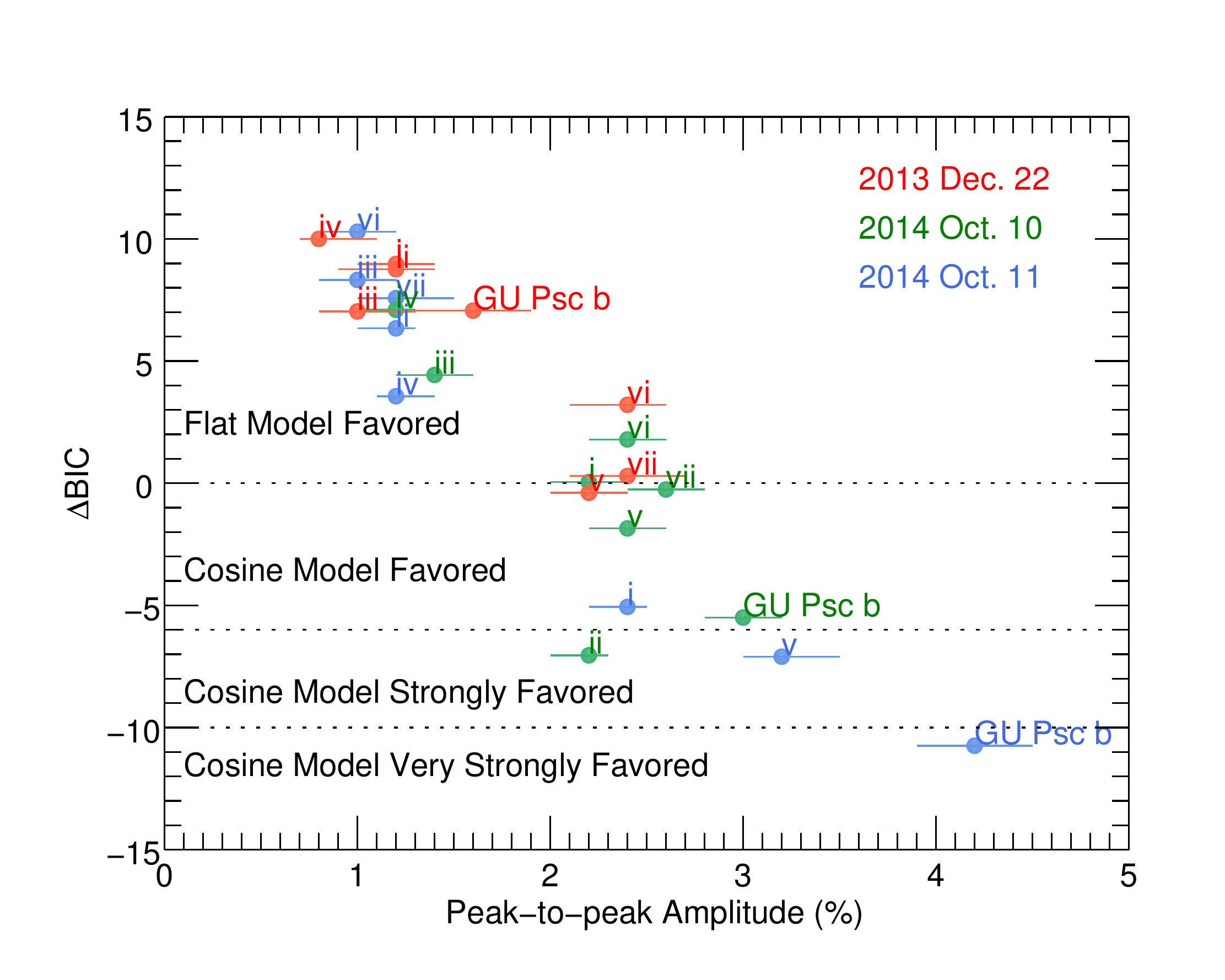}
\caption{Result of the BIC analysis, which compares the best cosine model and a flat model for GU~Psc~b and the 7 comparison stars, for the three epochs versus the peak-to-peak amplitude of the cosine model considered. The $\Delta$BIC of GU~Psc~b on the third epoch indicates that the cosine model is below $-10$, which means it is very strongly favored over the flat model, according to the \citet{Kass:2012bb} scale.}
\label{fig:deltaBIC}
\end{center}
\end{figure*}

The maximum likelihood for two models was computed on the light curves of  GU~Psc~b and the comparison stars in order to quantitatively assess the significance of the detection for GU~Psc~b at the third epoch. Model 1, the cosine model, used Equation \ref{eq:model} with $A$ fixed to the median value from the MCMC analysis (7 degrees of freedom, DOF).  Model 2 had $A$, $P$ and $t_0$ fixed to zero, which is equivalent to fitting a flat line  (a non-variable model) to the data simultaneously with principle components (5 DOF). The maximum likelihood was found using the L-BFGS-B code \citep{Zhu1997}. This code is a limited memory, quasi-Newton method that approximates the Broyden-Fletcher-Goldfarb-Shanno algorithm \citep{Press:1992vz}. The Bayesian Information Criterion (BIC) was then computed for each star and each epoch with $n$ data points with 7 DOF for model 1 and 5 DOF for model 2. The BIC penalizes model 1 for additional DOF. The difference between models 1 and 2 was computed. Figure \ref{fig:deltaBIC} shows the $\Delta$BIC$=$BIC$_{1}-$BIC$_{2}$ versus amplitude for all stars at each epoch. Lower relative values of the BIC indicate a preferred model. According to \citet{Kass:2012bb}, a $|\Delta$BIC$|$ between $6$ and $10$ indicate that one model is ``strongly'' favored over the other,  while values above $10$ means the best model is "very strongly" favored over the other. By far, the greatest $|\Delta$BIC$|$ ($10.2$) is found for GU~Psc~b on the third epoch, which means the cosine (variable) model with a peak-to-peak amplitude of 4.2\% is strongly preferred over the flat line (non-variable) in that case. No other star/epoch shows a variability of similar amplitude at this significance level.  

\section{Discussion}
\label{sec:analdis}
The high-precision photometry observations presented here suggest that GU~Psc~b, a young planetary-mass companion at the cool-end of the L/T transition, shows tentative variability on one of the three epochs in which it was observed (2014 October 11). The MCMC analysis suggests a variability with peak-to-peak amplitude of $4\pm1$\% on a timescale of \env6\,hr. If confirmed, this would be among the first variability measurements of a planetary-mass companion, GU~Psc~b being a rare example that is amenable to such high-precision measurements. The observed variability patterns (amplitude, timescale, secular variability) of GU~Psc~b appear consistent with those observed for brown dwarfs and planetary-mass objects. In the MCMC analysis, at least one of the seven comparison stars, star \textrm{v}, seems to vary significantly at the same epoch, on a shorter timescale and with a slightly lower amplitude. According to the BIC criterion, in that case, the cosine model is also favored over the straight line. The comparison stars, being randomly drawn in the field, are in all likelihood late-G, K or early M stars. Their $i-z$, $i-J$, $z-J$, and $J-K$ colors, measured in WIRCam \citep{Naud2014} and SDSS data, are consistent with that. It is thus plausible that the observed variability for comparison star v is also due to stellar variability (see, e.g.,  \citealt{McQuillan2014}).

Based on their \textit{Spitzer} survey of 44 field and young brown dwarfs, \citet{Metchev2015} confirmed that variability is widespread among L and T dwarfs, supporting the findings of \citet{Buenzli2014}. They found that 31\% of their single T0--T8 were variable with peak-to-peak amplitude between 0.2\% and 4.6\%, hence the amplitude suggested here for GU~Psc~b is expected if the mechanism at play in brown dwarfs extends to the planetary-mass regime. \citet{Radigan2014a} and \citet{Radigan2014b} argue that strong variability is even more prevalent for the L/T transition brown dwarfs, at the limit of which GU~Psc~b lies. 

It is not clear if young objects (often planetary-mass companions) display a similar variability to single field brown dwarfs. \citet{Metchev2015} suggested that in the L dwarf regime, the variable low-gravity Ls could have a higher variability amplitude than the variable field L dwarfs. In accordance with this finding, most of the field L dwarfs studied in the $J$ band by \citet{Radigan2014a} are not variable above the \env2\% level. In contrast, the young, late-L, free-floating planetary-mass object PSO~318.5-22 was discovered to be variable with an amplitude of 7--10\% in the $J$ band \citep{Biller2015}. Another young planetary-mass companion, 2M~1207~b, a mid= to late-L-type member of the TW Hya Association (\env 8\,Myr; \citealp{Gizis2002,Chauvin2004}), was also recently found to be variable in the near-infrared with an amplitude of 1.36\% in the HST F125W (similar to $J$ band) and 0.78\% in the F160W filters (similar to $H$ band; \citealp{Zhou2016}), as was the dusty L6 dwarf W0047, which was found to be variable with a peak-to-peak amplitude of 8\% with the WFC3ʼs G141 grism, which covers 1.075--1.7$\micron$ \citep{Lew2016}. 

In the T-dwarf regime, \citet{Metchev2015} did not find an increased variability frequency or variability amplitude for young T dwarfs. They studied in the mid-infrared two planetary-mass companions that bear some similarities to GU~Psc~b. The T2.5  HN~Peg~B, near the deuterium-burning limit, was found to be variable with amplitudes of 0.77\% and 1.1\% in the $[3.4]$ and $[4.5]$ filters, respectively, while the late T dwarf Ross 458 (AB) c showed no variability above 1.4\% and 0.7\% in the same filters. Both companions are somewhat older than GU~Psc~b but do not have ages that are very well constrained (HN~Peg~B, 300 $\pm$ 200 Myr; \citealp{Luhman2007,Leggett2008}, Ross~458~(AB)~c, 150--800\,Myr, \citealp{Burgasser2010_Ross}). It could, however, be expected that the variability in the mid-infrared is smaller in amplitude than the variability in $J$ band, where clouds of different temperatures and the atmosphere are expected to show the greatest contrast \citep{Ackerman2001,Marley2002}. For example, with simultaneous near-infrared and \textit{Spitzer} mid-infrared observations, \citet{Yang2016} showed that the amplitude in $J$ band is two to three times higher than in the \textit{Spitzer} wavelength for SIMP~0136+0933. This could thus suggest that HN~Peg~B would show variability in the $J$ band with an amplitude similar to that measured here for GU Psc b. The fact that the T2.5 SIMP~0136+0933 has probably itself an age that is similar to GU~Psc~b (around \env200\,Myr according to its plausible membership in Carina-Near) is also interesting, since it is one of the most variable objects known.

No high-significance periodic variability was detected for the first two epochs, which were taken approximately 10 months and 1 day before the third epoch. Large-scale evolution of weather patterns are suspected to cause long-term changes in the light curve, which may explain these findings. Most brown dwarfs monitored to date show an evolution of their light curve. The 6\,yr-monitoring of SIMP~0136 showed that its peak-to-peak amplitude varied between \env2$\%$ to more than 10$\%$ \citep{Metchev2013,Croll2016}. Observations of 2MASS~J2139 also suggest an evolution of its light curve over periods of several weeks \citep{Radigan2012}. Neptune, in the solar system, also shows a secular variability \citep{Simon2016}. The fact that the variability observed on 2014 October 11 is not seen on 2014 October 10 with the same significance is also something that can be expected. \citet{Metchev2015} found that many of their variable brown dwarfs showed an evolution of their light curve over timescales of hours only. The T1 Luhman~16B also displays an important evolution from night to night in the 12-day monitoring made with the TRAPPIST telescope \citep{Gillon2013}.

The likely variability detected here could be explained by rotation bringing features in and out of view. Brown dwarfs are known to have relatively short rotation periods, between 2 and \env20\,hr \citep{Metchev2015}. The \citet{Metchev2015} survey measured a period of 18\,hr for HN~Peg~B. GU~Psc~b could have a longer period because it is younger and thus has an inflated radius. For giant exoplanets, rotation periods are still largely unknown. Recently, near-infrared high-resolution spectroscopic observations of \citet{Snellen2014} allowed us to measure an equatorial rotation velocity of $V_{\rm{spin}}=25$\,\kms\ for the 7\,\MJ\ exoplanet $\beta$~Pictoris~b. They assumed that it has a 1.65 Jupiter radius, given that it is a member of the young association $\beta$~Pictoris ($<25$\,Myr, \citealp{Bonnefoy2013,Currie2013,Binks2014, Malo2014b}). They estimate a rotation period of about 8\,hr. In another young association (TW Hya), the young planetary-mass companion 2M1207b was recently observed to vary with a period of 10.7\,hr \citep{Zhou2016} and the free-floating planetary-mass object PSO~318 allowed to constrain its period to $>$5\,hr \citep{Biller2015}. A period greater than \env6\,hr for GU~Psc~b would thus not be surprising. 

Longer observations will be needed to confirm the variability of GU~Psc~b and its periodic nature, and to eventually better constrain the rotation period. GU~Psc~b is a prime target for long-term high-precision photometry observation on 8 m class telescopes, \textit{Spitzer} and JWST. As GU~Psc~b seems to show variability in the $J$ band, it would also be interesting to search for variability in other near-infrared or mid-infrared bands, as simultaneous observations in different bands allow to probe different layers of the atmosphere \citep{Buenzli2012,Apai2013,Biller2013b,Yang2016}. GU~Psc~b has a $W2=15.41$ \citep{Naud2014}, so according to the typical performances achieved with \textit{Spitzer} in \citeauthor{Metchev2015} (2015; see Figure 7), this instrument would be able to detect a variability amplitude down to about $3-4$\% in the $[4.5]$ band.

\subsection*{Acknowledgments}
The authors would like to thank the CFHT staff for their precious help throughout this project. They would also like to thank the anonymous referee and AAS statistics editors for constructive comments and suggestions that improved the overall quality of the paper. This work was financially supported by the Natural Sciences and Engineering Research Council (NSERC) of Canada and the Fond de Recherche Qu\'{e}b\'{e}cois - Nature et Technologie (FRQNT; Qu\'{e}bec). 
Based on observations obtained at CFHT with WIRCam.
This publication makes use of data products from the Two Micron All Sky Survey, which is a joint project of the University of Massachusetts and the Infrared Processing and Analysis Center, and funded by the National Aeronautics and Space Administration and the National Science Foundation, of the NASA's Astrophysics Data System Bibliographic Services, SIMBAD database, the VizieR catalog access tool and the SIMBAD database operated at CDS, Strasbourg, France.

\software{I'iwi v2.1.200; IDL MPFIT2DPEAK, APER, PCA}

\section*{Appendix: Additional figures}

\begin{figure*}[htbp]
\begin{center}
\includegraphics[width=13.5cm]{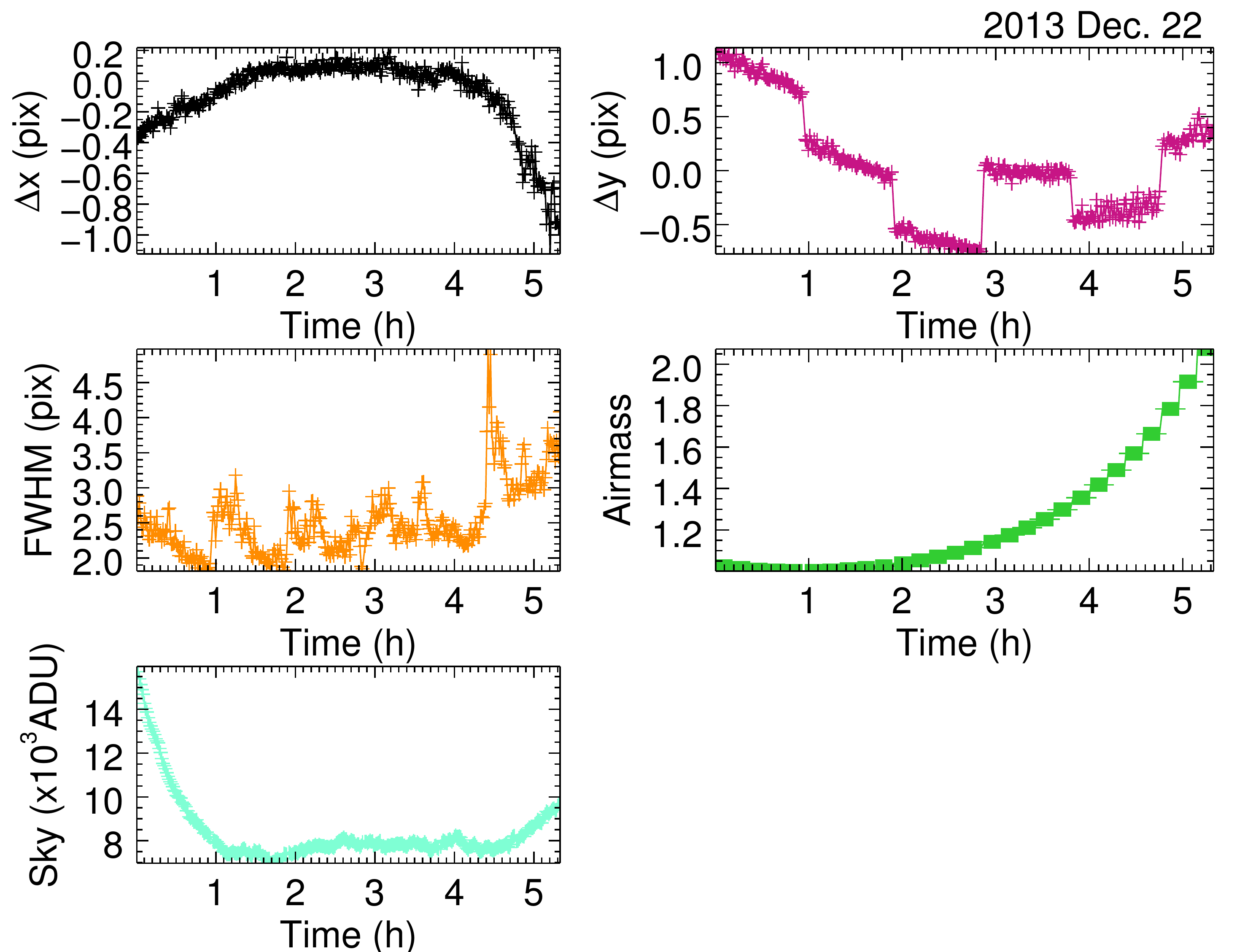}
\includegraphics[width=13.5cm]{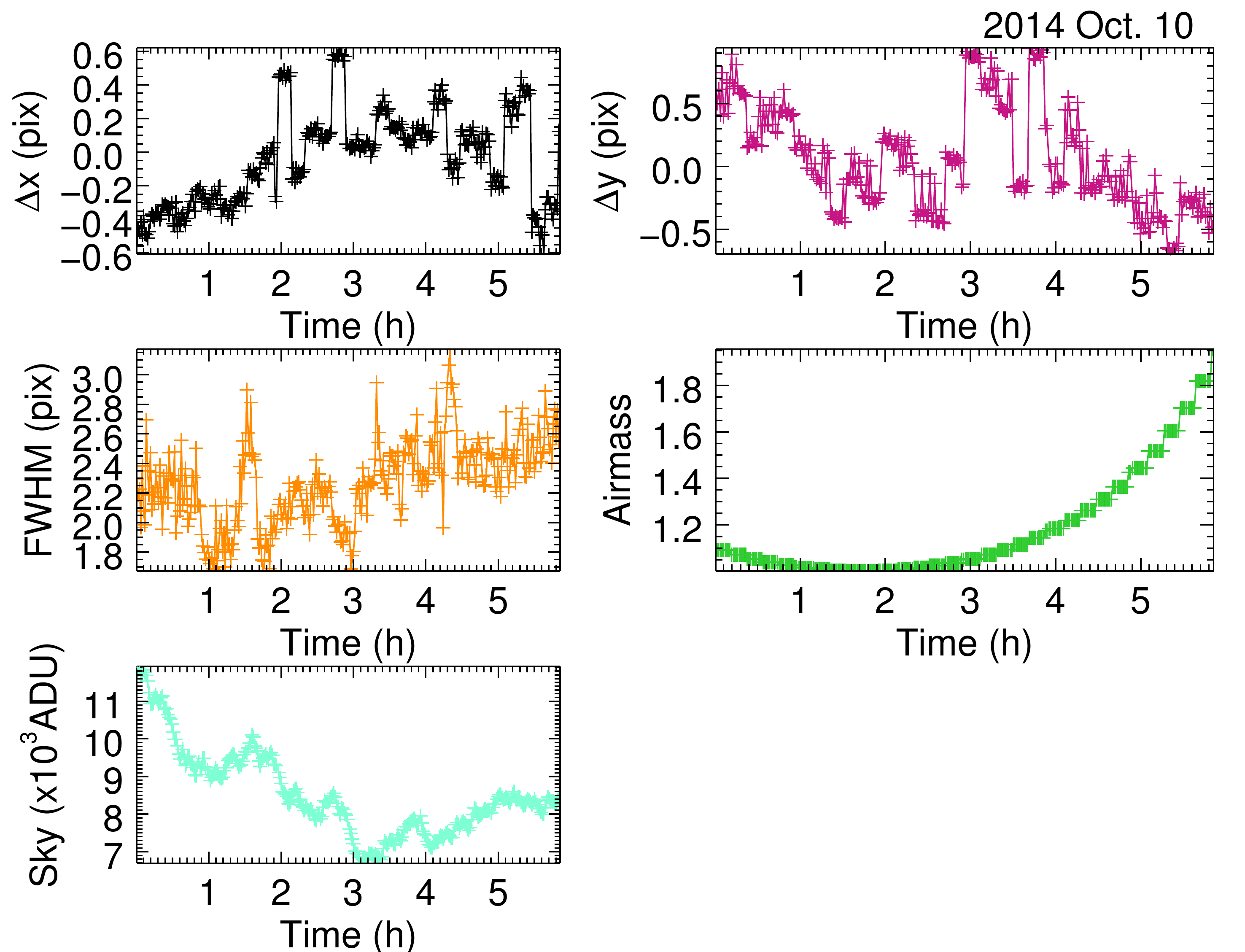}
\caption{Variation of different external parameters for 2013 December 22 (top) and 2014 October 10 (bottom). The variation of the x and y positions on the chip (median of all stars), FWHM, airmass, and sky level (ADU) are shown.}
\label{fig:extparam2}
\end{center}
\end{figure*}

\begin{figure*}[htbp]
\begin{center}
\includegraphics[width=10.5cm]{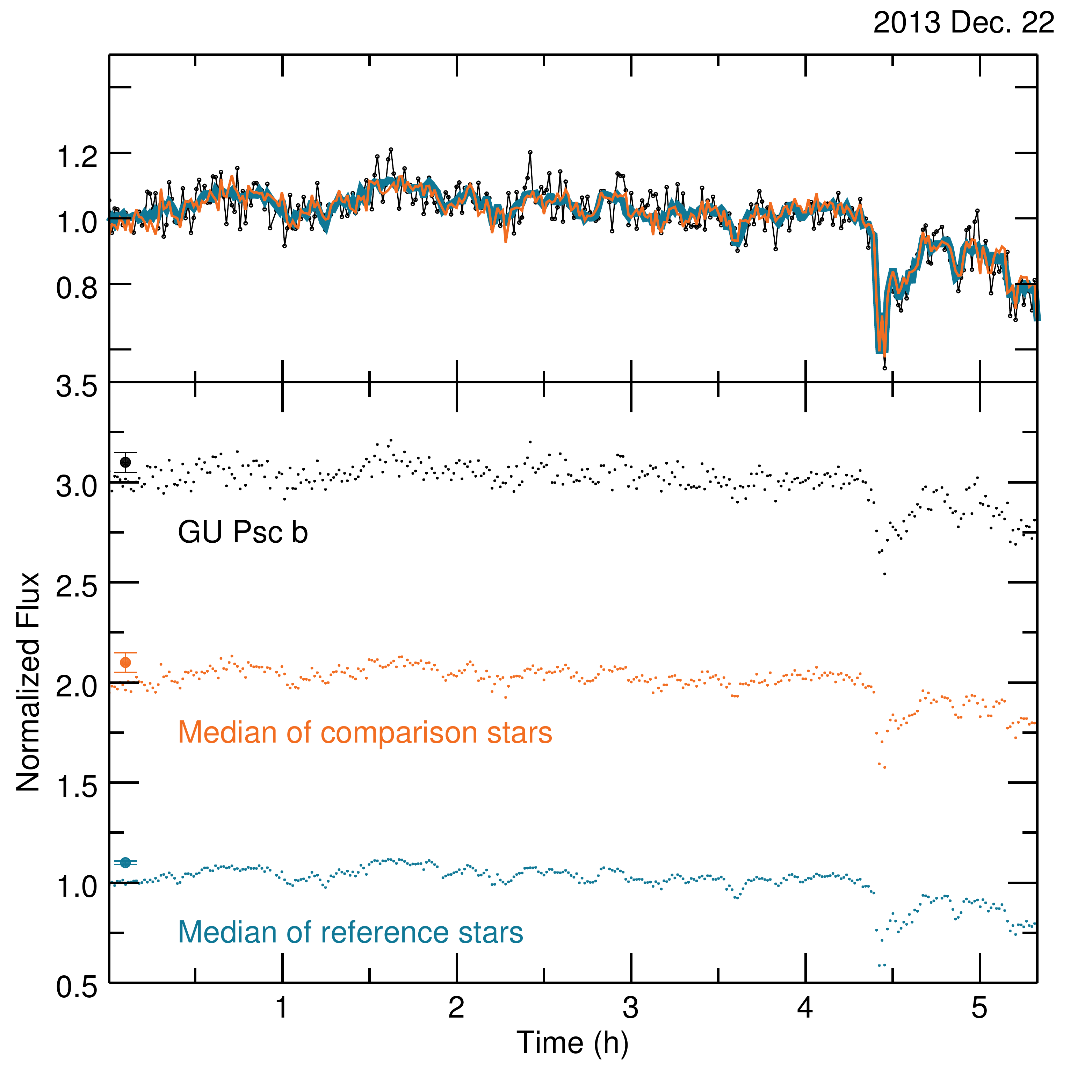}
\includegraphics[width=10.5cm]{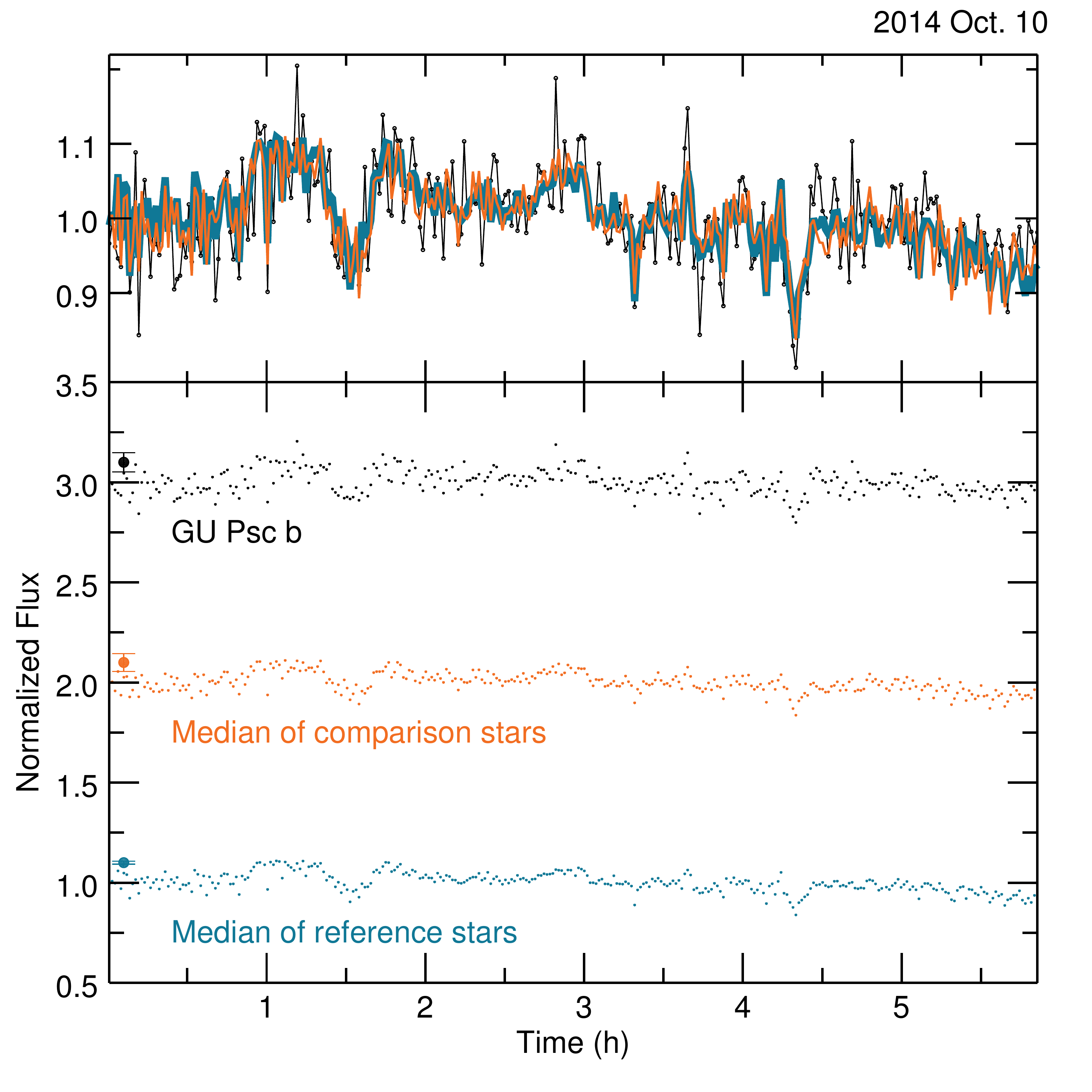}
\caption{Raw normalized light curves for GU~Psc~b (top curve), the median of comparison stars (middle) and the median of all reference stars (bottom), for 2013 December 22 (top) and 2014 October 10 (bottom). GU~Psc~b and comparison star median curves have been offset for clarity. For each date, a superposition of the three curves is also shown above the three curves. GU~Psc~b is the black curve, with dot symbols, the median of the reference stars is the thick dark cyan line, and the median of the comparison stars is the thin orange pale line.}
\label{fig:raw2}
\end{center}
\end{figure*}

\begin{figure*}[htbp]
\begin{center}
\includegraphics[width=14.5cm]{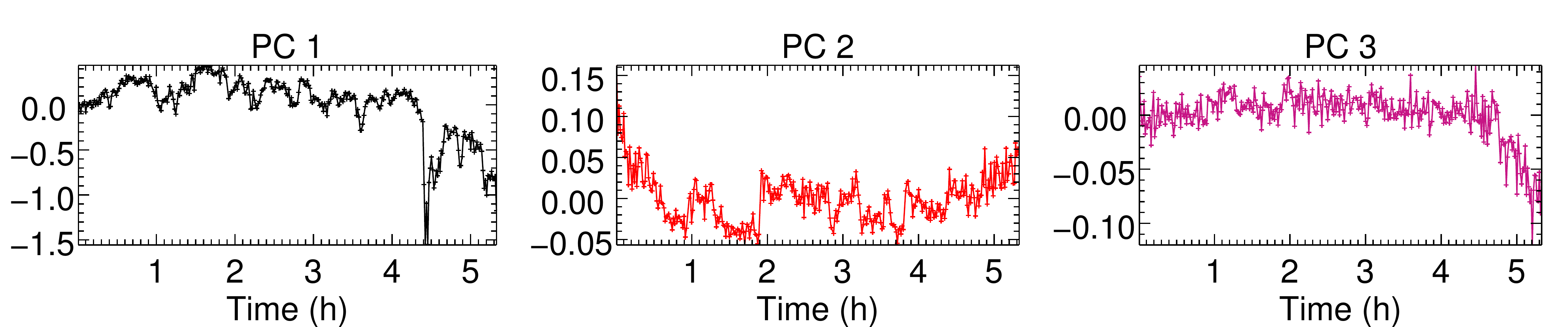}
\includegraphics[width=14.5cm]{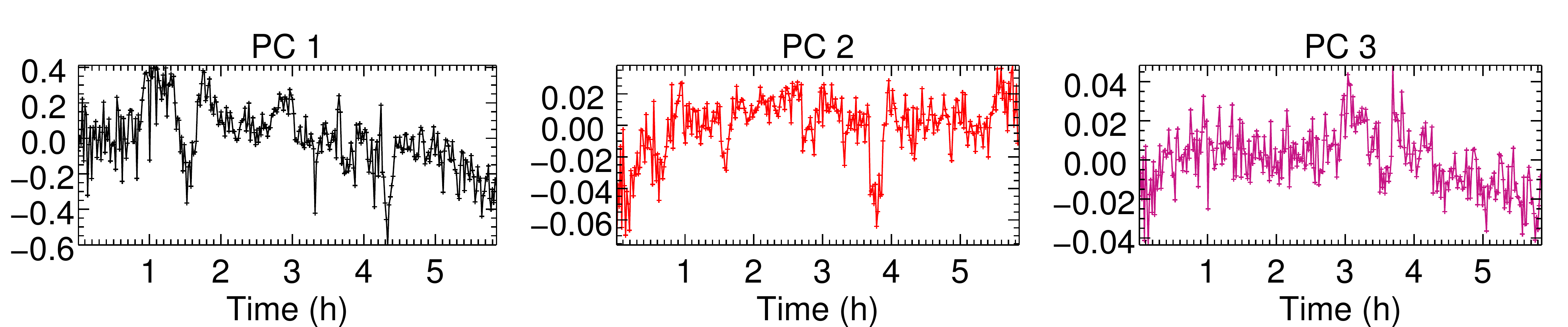}
\caption{Most important principal components retained for for 2013 December 22 (top) and 2014 October 10 (bottom).}
\label{fig:PC2}
\end{center}
\end{figure*}

Raw light curves, external parameter evolution, and retained principal components for the first and second epochs are shown in the Figures \ref{fig:extparam2}, \ref{fig:raw2} and \ref{fig:PC2}. 

\begin{thebibliography}{}
\expandafter\ifx\csname natexlab\endcsname\relax\def\natexlab#1{#1}\fi

\bibitem[{Ackerman \& Marley(2001)}]{Ackerman2001}
Ackerman, A.~S., \& Marley, M.~S. 2001, ApJ, 556, 872

\bibitem[{Apai {et~al.}(2013)Apai, Radigan, Buenzli, Burrows, Reid, \&
  Jayawardhana}]{Apai2013}
Apai, D., Radigan, J., Buenzli, E., {et~al.} 2013, ApJ, 768, 121

\bibitem[{Apai {et~al.}(2016)Apai, Kasper, Skemer, Hanson, Lagrange, Biller,
  Bonnefoy, Buenzli, \& VIGAN}]{Apai2016}
Apai, D., Kasper, M., Skemer, A., {et~al.} 2016, ApJ, 820, 40

\bibitem[{Artigau {et~al.}(2009)Artigau, Bouchard, Doyon, \&
  Lafreni{\`e}re}]{Artigau2009}
Artigau, {\'E}., Bouchard, S., Doyon, R., \& Lafreni{\`e}re, D. 2009, ApJ, 701,
  1534

\bibitem[{Bell {et~al.}(2015)Bell, Mamajek, \& Naylor}]{Bell2015}
Bell, C. P.~M., Mamajek, E.~E., \& Naylor, T. 2015, MNRAS, 454, 593

\bibitem[{Biller {et~al.}(2013)Biller, Crossfield, Mancini, Ciceri, Southworth,
  Kopytova, Bonnefoy, Deacon, Schlieder, Buenzli, Brandner, Allard, Homeier,
  Freytag, Bailer-Jones, Greiner, Henning, \& Goldman}]{Biller2013b}
Biller, B.~A., Crossfield, I. J.~M., Mancini, L., {et~al.} 2013, ApJL, 778, L10

\bibitem[{Biller {et~al.}(2015)Biller, Vos, Bonavita, Buenzli, Baxter,
  Crossfield, Allers, Liu, Bonnefoy, Deacon, Brandner, Schlieder, Dupuy,
  Kopytova, Manjavacas, Allard, Homeier, \& Henning}]{Biller2015}
Biller, B.~A., Vos, J., Bonavita, M., {et~al.} 2015, ApJL, 813, L23

\bibitem[{Binks \& Jeffries(2014)}]{Binks2014}
Binks, A.~S., \& Jeffries, R.~D. 2014, MNRAS, 438, L11

\bibitem[{Bonnefoy {et~al.}(2013)Bonnefoy, Boccaletti, Lagrange, Allard,
  Mordasini, Beust, Chauvin, Girard, Homeier, Apai, Lacour, \&
  Rouan}]{Bonnefoy2013}
Bonnefoy, M., Boccaletti, A., Lagrange, A.~M., {et~al.} 2013, A{\&}A, 555, A107

\bibitem[{Brooks \& Gelman(1998)}]{Brooks1998}
Brooks, S.~P., \& Gelman, A. 1998, Journal of computational and graphical, 7, 4, 434–455
  {\ldots}

\bibitem[{Buenzli {et~al.}(2014)Buenzli, Apai, Radigan, Reid, \&
  Flateau}]{Buenzli2014}
Buenzli, E., Apai, D., Radigan, J., Reid, I.~N., \& Flateau, D. 2014, ApJ, 782,
  77

\bibitem[{Buenzli {et~al.}(2015)Buenzli, Saumon, Marley, Apai, Radigan, Bedin,
  Reid, \& Morley}]{Buenzli2015}
Buenzli, E., Saumon, D., Marley, M.~S., {et~al.} 2015, ApJ, 798, 127

\bibitem[{Buenzli {et~al.}(2012)Buenzli, Apai, Morley, Flateau, Showman,
  Burrows, Marley, Lewis, \& Reid}]{Buenzli2012}
Buenzli, E., Apai, D., Morley, C.~V., {et~al.} 2012, ApJ, 760, L31

\bibitem[{Burgasser {et~al.}(2010)Burgasser, Simcoe, Bochanski, Saumon,
  Mamajek, Cushing, Marley, McMurtry, Pipher, \& Forrest}]{Burgasser2010_Ross}
Burgasser, A.~J., Simcoe, R.~A., Bochanski, J.~J., {et~al.} 2010, ApJ, 725,
  1405

\bibitem[{Burgasser {et~al.}(2014)Burgasser, Gillon, Faherty, Radigan, Triaud,
  Plavchan, Street, Jehin, Delrez, \& Opitom}]{Burgasser2014}
Burgasser, A.~J., Gillon, M., Faherty, J.~K., {et~al.} 2014, ApJ, 785, 48

\bibitem[{Chauvin {et~al.}(2004)Chauvin, Lagrange, Dumas, Zuckerman, Mouillet,
  Song, Beuzit, \& Lowrance}]{Chauvin2004}
Chauvin, G., Lagrange, A.~M., Dumas, C., {et~al.} 2004, A{\&}A, 425, L29

\bibitem[{Croll {et~al.}(2016)Croll, Muirhead, Lichtman, Han, Dalba, \&
  Radigan}]{Croll2016}
Croll, B., Muirhead, P.~S., Lichtman, J., {et~al.} 2016, 1609.03587

\bibitem[{Crossfield {et~al.}(2014)Crossfield, Biller, Schlieder, Deacon,
  Bonnefoy, Homeier, Allard, Buenzli, Henning, Brandner, Goldman, \&
  Kopytova}]{Crossfield2014}
Crossfield, I. J.~M., Biller, B.~A., Schlieder, J.~E., {et~al.} 2014, Nature,
  505, 654

\bibitem[{Currie {et~al.}(2013)Currie, Burrows, Madhusudhan, Fukagawa, Girard,
  Dawson, Murray-Clay, Kenyon, Kuchner, Matsumura, Jayawardhana, Chambers, \&
  Bromley}]{Currie2013}
Currie, T., Burrows, A.~S., Madhusudhan, N., {et~al.} 2013, ApJ, 776, 15

\bibitem[{Cushing {et~al.}(2016)Cushing, Hardegree-Ullman, Trucks, Morley,
  Gizis, Marley, Fortney, Kirkpatrick, Gelino, Mace, \& Carey}]{Cushing2016}
Cushing, M.~C., Hardegree-Ullman, K.~K., Trucks, J.~L., {et~al.} 2016, ApJ,
  823, 152

\bibitem[{Devost {et~al.}(2010)Devost, Albert, Teeple, \& Croll}]{Devost2010}
Devost, D., Albert, L., Teeple, D., \& Croll, B. 2010, in SPIE Astronomical
  Telescopes and Instrumentation: Observational Frontiers of Astronomy for the
  New Decade, ed. D.~R. Silva, A.~B. Peck, \& B.~T. Soifer (SPIE),
  77372D

\bibitem[{Gagn{\'e} {et~al.}(2017)Gagn{\'e}, Faherty, Burgasser, Artigau,
  Bouchard, Albert, Lafreni{\`e}re, Doyon, \& Bardalez~Gagliuffi}]{Gagne2017}
Gagn{\'e}, J., Faherty, J.~K., Burgasser, A.~J., {et~al.} 2017, ApJ, 841, L1

\bibitem[{Gelman \& Rubin(1992)}]{Gelman1992}
Gelman, A., \& Rubin, D.~B. 1992, Statistical science

\bibitem[{Gillon {et~al.}(2013)Gillon, Triaud, Jehin, Delrez, Opitom, Magain,
  Lendl, \& Queloz}]{Gillon2013}
Gillon, M., Triaud, A. H. M.~J., Jehin, E., {et~al.} 2013, A{\&}A, 555, L5

\bibitem[{Girardin {et~al.}(2013)Girardin, Artigau, \& Doyon}]{Girardin2013}
Girardin, F., Artigau, {\'E}., \& Doyon, R. 2013, ApJ, 767, 61

\bibitem[{Gizis(2002)}]{Gizis2002}
Gizis, J.~E. 2002, ApJ, 575, 484

\bibitem[{Gregory(2010)}]{Gregory2011}
Gregory, P.~C. 2010, MNRAS, 410, 94

\bibitem[{Jolliffe(2002)}]{Jolliffe2002}
Jolliffe, I.~T. 2002, {Principal Component Analysis}, second edition edn.,
  Springer Series in Statistics (New York: Springer-Verlag)

\bibitem[{Kass \& Raftery(1995)}]{Kass:2012bb}
Kass, R.~E., \& Raftery, A.~E. 1995, J. Am. Stat. Assoc., 90, 430, 773

\bibitem[{Kostov \& Apai(2012)}]{Kostov2013}
Kostov, V., \& Apai, D. 2012, ApJ, 762, 47

\bibitem[{Leggett {et~al.}(2008)Leggett, Saumon, Albert, Cushing, Liu, Luhman,
  Marley, Kirkpatrick, Roellig, \& Allers}]{Leggett2008}
Leggett, S.~K., Saumon, D., Albert, L., {et~al.} 2008, ApJ, 682, 1256

\bibitem[{Lew {et~al.}(2016)Lew, Apai, Zhou, Schneider, Burgasser, Karalidi,
  Yang, Marley, Cowan, Bedin, Metchev, Radigan, \& Lowrance}]{Lew2016}
Lew, B. W.~P., Apai, D., Zhou, Y., {et~al.} 2016, ApJL, 829, L32

\bibitem[{Luhman {et~al.}(2007)Luhman, Patten, Marengo, Schuster, Hora, Ellis,
  Stauffer, Sonnett, Winston, Gutermuth, Megeath, Backman, Henry, Werner, \&
  Fazio}]{Luhman2007}
Luhman, K.~L., Patten, B.~M., Marengo, M., {et~al.} 2007, ApJ, 654, 570

\bibitem[{Malo {et~al.}(2014)Malo, Doyon, Feiden, Albert, Lafreni{\`e}re,
  Artigau, Gagn{\'e}, \& Riedel}]{Malo2014b}
Malo, L., Doyon, R., Feiden, G.~A., {et~al.} 2014, ApJ, 792, 37

\bibitem[{Marley {et~al.}(2002)Marley, Seager, Saumon, Lodders, Ackerman,
  Freedman, \& Fan}]{Marley2002}
Marley, M.~S., Seager, S., Saumon, D., {et~al.} 2002, ApJ, 568, 335

\bibitem[{McQuillan {et~al.}(2014)McQuillan, Mazeh, \& Aigrain}]{McQuillan2014}
McQuillan, A., Mazeh, T., \& Aigrain, S. 2014, ApJS, 211, 24

\bibitem[{Metchev {et~al.}(2013)Metchev, Apai, Radigan, Artigau, Heinze,
  Helling, Homeier, Littlefair, Morley, Skemer, \& Stark}]{Metchev2013}
Metchev, S., Apai, D., Radigan, J., {et~al.} 2013, AN, 334, 40

\bibitem[{Metchev {et~al.}(2015)Metchev, Heinze, Apai, Flateau, Radigan,
  Burgasser, Marley, Artigau, Plavchan, \& Goldman}]{Metchev2015}
Metchev, S.~A., Heinze, A., Apai, D., {et~al.} 2015, ApJ, 799, 154

\bibitem[{Morley {et~al.}(2014)Morley, Marley, Fortney, \& Lupu}]{Morley2014}
Morley, C.~V., Marley, M.~S., Fortney, J.~J., \& Lupu, R. 2014, ApJL, 789, L14

\bibitem[{Naud {et~al.}(2014)Naud, Artigau, Malo, Albert, Doyon,
  Lafreni{\`e}re, Gagn{\'e}, Saumon, Morley, Allard, Homeier, Beichman, Gelino,
  \& Boucher}]{Naud2014}
Naud, M.-E., Artigau, {\'E}., Malo, L., {et~al.} 2014, ApJ, 787, 5

\bibitem[{Naud {et~al.}(2017)Naud, Artigau, Doyon, Malo, Gagn{\'e}, Lafreni{\`e}re, Wolf, \& Magnier}]{Naud2017a}
Naud, M.-E., Artigau, {\'E}., Doyon, R., {et~al.} 2017, AJ, 154, 3

\bibitem[{Press {et~al.}(1992)Press, Teukolsky, Vetterling, \&
  Flannery}]{Press:1992vz}
Press, W.~H., Teukolsky, S.~A., Vetterling, W.~T., \& Flannery, B.~P. 1992,
  Cambridge: University Press, |c1992, 2nd ed.

\bibitem[{Puget {et~al.}(2004)Puget, Stadler, Doyon, Gigan, Thibault, Luppino,
  Barrick, Benedict, Forveille, Rambold, Thomas, Vermeulen, Ward, Beuzit,
  Feautrier, Magnard, Mella, Preis, Vallee, Wang, Lin, Hall, \&
  Hodapp}]{Puget2004}
Puget, P., Stadler, E., Doyon, R., {et~al.} 2004, Proc. SPIE, 5494, 978

\bibitem[{Radigan(2014)}]{Radigan2014b}
Radigan, J. 2014, ApJ, 797, 120

\bibitem[{Radigan {et~al.}(2012)Radigan, Jayawardhana, Lafreni{\`e}re, Artigau,
  Marley, \& Saumon}]{Radigan2012}
Radigan, J., Jayawardhana, R., Lafreni{\`e}re, D., {et~al.} 2012, ApJ, 750, 105

\bibitem[{Radigan {et~al.}(2014)Radigan, Lafreni{\`e}re, Jayawardhana, \&
  Artigau}]{Radigan2014a}
Radigan, J., Lafreni{\`e}re, D., Jayawardhana, R., \& Artigau, {\'E}. 2014,
  ApJ, 793, 75

\bibitem[{Robinson \& Marley(2014)}]{Robinson2014}
Robinson, T.~D., \& Marley, M.~S. 2014, ApJ, 785, 158

\bibitem[{Rowe {et~al.}(2014)Rowe, Bryson, Marcy, Lissauer, Jontof-Hutter,
  Mullally, Gilliland, Issacson, Ford, Howell, Borucki, Haas, Huber, Steffen,
  Thompson, Quintana, Barclay, Still, Fortney, Gautier, Hunter, Caldwell,
  Ciardi, Devore, Cochran, Jenkins, Agol, Carter, \& Geary}]{Rowe2014}
Rowe, J.~F., Bryson, S.~T., Marcy, G.~W., {et~al.} 2014, ApJ, 784, 45

\bibitem[{Showman \& Kaspi(2013)}]{ShowmanKaspi2013}
Showman, A.~P., \& Kaspi, Y. 2013, ApJ, 776, 85

\bibitem[{Simon {et~al.}(2016)Simon, Rowe, Gaulme, Hammel, Casewell, Fortney,
  Gizis, Lissauer, Morales-Juberias, Orton, Wong, \& Marley}]{Simon2016}
Simon, A.~A., Rowe, J.~F., Gaulme, P., {et~al.} 2016, ApJ, 817, 162

\bibitem[{Snellen {et~al.}(2014)Snellen, Brandl, de~Kok, Brogi, Birkby, \&
  Schwarz}]{Snellen2014}
Snellen, I. A.~G., Brandl, B.~R., de~Kok, R.~J., {et~al.} 2014, Nature, 509, 63

\bibitem[{Wilson {et~al.}(2014)Wilson, Rajan, \& Patience}]{Wilson2014}
Wilson, P.~A., Rajan, A., \& Patience, J. 2014, A{\&}A, 566, A111

\bibitem[{Yang {et~al.}(2016)Yang, Apai, Marley, Karalidi, Flateau, Showman,
  Metchev, Buenzli, Radigan, Artigau, Lowrance, \& Burgasser}]{Yang2016}
Yang, H., Apai, D., Marley, M.~S., {et~al.} 2016, ApJ, 826, 8

\bibitem[{Zhou {et~al.}(2016)Zhou, Apai, Schneider, Marley, \&
  Showman}]{Zhou2016}
Zhou, Y., Apai, D., Schneider, G.~H., Marley, M.~S., \& Showman, A.~P. 2016,
  ApJ, 818, 176

\bibitem[{Zhu {et~al.}(1997)Zhu, Byrd, Lu, \& Nocedal}]{Zhu1997}
Zhu, C., Byrd, R.~H., Lu, P., \& Nocedal, J. 1997, ACM Transactions on
  Mathematical Software (TOMS), 23, 550

\end{thebibliography}

\end{document}